\documentclass[a4paper,11pt]{article}
\pdfoutput=1 

\usepackage{jheppub} 
                     
\usepackage[T1]{fontenc} 
\usepackage[dvipsnames]{xcolor}
\usepackage{subfigure}
\usepackage{braket}
\usepackage{bbm}
\usepackage[normalem]{ulem}
\usepackage{hyperref}

\def\Tr{\mathop{\mbox{\normalfont Tr}}\nolimits}
\def\chartheta#1#2{\Theta\!\left[\begin{smallmatrix}#1\\#2\end{smallmatrix}\right]\!}

\begin{document}

\title{\boldmath Multi-charged moments of two intervals
in conformal field theory
}

\vspace{.5cm}

\author{Filiberto Ares$^1$, Pasquale Calabrese$^{1,2}$, Giuseppe Di Giulio$^3$ and Sara Murciano$^1$}
\affiliation{$^1$SISSA and INFN Sezione di Trieste, via Bonomea 265, 34136 Trieste, Italy.}
\affiliation{$^{2}$International Centre for Theoretical Physics (ICTP), Strada Costiera 11, 34151 Trieste, Italy.}
\affiliation{$^{3}$Institute for Theoretical Physics and Astrophysics and W\"urzburg-Dresden Cluster of Excellence ct.qmat, Julius-Maximilians-Universit\"at W\"urzburg, Am Hubland, 97074 W\"{u}rzburg, Germany.}

\emailAdd{smurcian@sissa.it}

\vspace{.5cm}

\abstract{ 
We study the multi-charged moments for two disjoint intervals in the 
ground state of two $1+1$ dimensional CFTs with central charge $c=1$ and global 
$U(1)$ symmetry: the massless Dirac field theory and the compact boson (Luttinger liquid).
For this purpose, we compute the partition function on the higher genus Riemann surface arising from the replica method in the presence of 
background magnetic fluxes between the sheets of the surface. 
We consider the general situation in which the fluxes generate different twisted boundary conditions at each branch point. 
The obtained multi-charged moments allow us to derive the symmetry resolution of the R\'enyi entanglement entropies and the mutual information
for non complementary bipartitions. We check our findings against exact numerical results for the 
tight-binding model, which is a lattice realisation of the massless Dirac theory.}

\maketitle

\section{Introduction}\label{sec:intro}
As Schr\"odinger already recognised one century ago, entanglement is at the core of quantum mechanics. 
Nowadays it turns out to be the fundamental notion behind many quantum phenomena, from quantum algorithms~\cite{nc-10} to gravity~\cite{Rangamani, nrt-09}, passing 
by critical phenomena and topological phases of matter~\cite{intro1, intro2, eisert-2010, intro3}, triggering unexpected 
connections between apparently far branches of physics.  At the center of all these ideas, we find the (R\'enyi) entanglement entropies which
are powerful entanglement measures that provide fundamental insights about the investigated system or theory.  
They are defined as follows.
Let us consider an extended quantum system in a pure state $|\Psi\rangle$ and a spatial bipartition into $A$ and $B$. 
The subsystem $A$ is described by the reduced density matrix $\rho_A=\mathrm{Tr}_B |\Psi\rangle\langle \Psi|$ and the associated R\'enyi entropies are 
given by the moments of $\rho_A$ as
\begin{equation}
\label{eq:Renyidef}
S_n^{A}\equiv \frac{1}{1-n}\ln \mathrm{Tr}[\rho_A^n],
\end{equation}
where we assume that $n$ is an integer number. After the analytic continuation to complex values of $n$, the limit $n \to 1$ of
Eq.~\eqref{eq:Renyidef} yields the von Neumann entanglement entropy
\begin{equation}
\label{eq:defEE}
S_1^A\equiv -\mathrm{Tr}[\rho_A\ln \rho_A].
\end{equation}
For bipartite systems in a pure state, the von Neumann and R\'enyi entropies 
can be used as measures of the entanglement shared between the two complementary parts.
One of the most interesting properties of entanglement entropies is that they 
are sensitive to criticality. In particular, for one-dimensional
gapless systems, if $A$ is a single interval, then the ground state 
entanglement entropy breaks the area law and is proportional to the 
central charge of the 1+1 dimensional CFT that describes the low-energy 
spectrum of the system~\cite{cw-94, hlw-94, cc-04, cc-09}. 

In the case considered in this work in which $A$ consists of two subsystems $A_1$ and 
$A_2$, i.e. $A=A_1 \cup A_2$,  the ground-state entanglement entropy  depends on the full operator content
of the CFT, encoding all the conformal data of the model~\cite{Caraglio, Furukawa, twist2, cct-11}. It is important to 
remark that, in this situation, the entanglement entropies quantify the 
entanglement between $A$ and $B$ but not between the two parts of $A$, for which 
one must resort to other entanglement measures such as negativity~\cite{neg-qft, neg-qft-2, ctt-13, Alba13, Coser3, Ares, Rockwood}. 
Nevertheless, from the entanglement entropies, it is possible to construct the following quantity, dubbed mutual information, 
\begin{equation}
\label{eq:defMI}
I^{A_1: A_2}\equiv S_1^{A_1}+S_1^{A_2}-S_1^{A},
\end{equation}
which is a measure of the total correlations between $A_1$ and $A_2$. The 
computation of two-interval R\'enyi entanglement entropies is a difficult problem,
even for minimal CFTs~\cite{Dupic, Ares}, as it boils down in general to determine the partition function of the 
theory on a higher genus $n$-sheeted Riemann surface~\cite{twist2, cct-11}. In fact, exact analytic expressions are 
only available for the free theories or special limits~\cite{Furukawa, twist2, cct-11, CFH, Alba, Alba2, twist3, CoserTagliacozzo, DeNobili, Coser2, rtc-18, HLM13, gkt-21, Casini,g-21,b-19,bds-20}. 
Moreover, the analytic continuation in $n$ to obtain Eq.~\eqref{eq:defEE} is still a challenging open issue.

In recent times, a question that has attracted much attention is 
how entanglement decomposes into the different symmetry sectors in 
the presence of global conserved charges~\cite{lr-14, goldstein, xavier}. Various reasons have motivated the interest in this problem.
The effect of symmetries on entanglement can be investigated experimentally~\cite{fis, Azses, Neven, Vitale} and, moreover, understanding how entanglement arises from the symmetry sectors is crucial to better grasp some quantum features, for example in 
non-equilibrium dynamics~\cite{fis}. Also at 
more practical level, it can help to speed-up the numerical algorithms
to simulate quantum many-body systems~\cite{xavier}. 
All that has been the breeding ground for a plethora 
of works that analyse the resolution of entanglement from different
perspectives: spin chains~\cite{lr-14, Vitale, Neven, riccarda, SREE2dG, goldstein2, MDC-19-CTM, ccgm-20, wv-03, bhd-18, bcd-19, byd-20, mrc-20, tr-19, ms-21, amc-22, jones-22, pvcc-22}, integrable quantum field theories~\cite{MDC-20, hcc-21, hcc-a-21, hc-20, chcc-a22}, CFTs~\cite{goldstein, xavier, goldstein1, crc-20, MBC-21, Chen-21, Capizzi-Cal-21, Hung-Wong-21, cdm-21, boncal-21, eim-d-21, Chen-22, Ghasemi-22,mt-22, ms-21}, holography~\cite{znm-20,wznm-21,znwm-22,bbcg-22},
out-of-equilibrium~\cite{pbc-21-1, pbc-21, fg-21, pbc-22, sh-22, chen-22-2} and disordered systems~\cite{trac-20, kusf-20, kusf-20b, kufs-21-1} or topological 
matter~\cite{clss-19, ms-20, Azses-Sela-20,ahn-20,ads-21,ore-21} to mention some of them.
In order to analyse entanglement in each symmetry
sector, quantities such as the symmetry-resolved entanglement entropy
~\cite{lr-14, goldstein, xavier} and the symmetry-resolved mutual information~\cite{pbc-21} have been proposed.
As shown in Ref.~\cite{goldstein}, symmetry-resolved entropies are intimately related to the charged moments of the 
reduced density matrix $\rho_A$, which were independently studied in holographic 
theories~\cite{Belin-Myers-13-HolChargedEnt, cms-13, cnn-16, d-16, d-17, ssr-17, shapourian-19}. 
Similarly to the moments of $\rho_A$, they can be interpreted as the partition function 
of the field theory on a Riemann surface, which is now coupled to an external magnetic flux. 
Partition functions with a
background gauge field have been also introduced as non-local order parameters to 
detect symmetry-protected topological phases in interacting fermionic systems~\cite{ssr1-17,ssr-16}. 

The symmetry resolution of entanglement in the two-interval case  has not been
much explored in CFT. Ref.~\cite{wznm-21} studies it at large central charge, in the context of holography, while, 
in Ref.~\cite{Chen-22}, the charged R\'enyi negativity is analysed for the complex free boson.
Here we take a different charge for each part of $A$, which leads to introduce the \textit{multi-charged} moments of $\rho_A$. 
This non-trivial generalisation of the charged moments, first considered in Ref.~\cite{pbc-21} in the context of quench dynamics, is the main 
subject of this work. In CFT, they correspond to the partition function on the $n$-sheeted Riemann surface, but with the insertion of a different 
magnetic flux across each subset (interval) of $A$.
We compute the multi-charged moments analytically for the ground state of two bidimensional CFTs with central charge $c=1$ and global $U(1)$ 
symmetry--- the massless Dirac field theory and the free compact boson---
generalising the expressions for the (neutral) R\'enyi entropies found in Refs.~\cite{CFH} and \cite{twist2} respectively. 
From the multi-charged moments, we 
derive the ground state symmetry-resolved entanglement entropy and mutual information
of two disjoint intervals.

The paper is organised as follows: in Sec.~\ref{sec:Definitions}, we define 
the symmetry-resolved entanglement and mutual information as well as the multi-charged moments, and 
we briefly describe the general approach to compute the latter in CFTs. 
We then move on to calculate the multi-charged moments for 
the ground state of the massless Dirac field theory in
Sec.~\ref{sec:FreeDirac} and of the free compact 
boson in Sec.~\ref{sec:CompactBoson}. In 
Sec.~\ref{sec:SymmetryResolution}, we apply the
previous results to obtain the symmetry-resolution of the mutual information 
in these theories. When possible, we benchmark the analytic expressions with exact numerical calculations for lattice models in the same universality class.  
We draw our conclusions in Sec.~\ref{sec:concl} and we include three appendices, with more details about the analytical and numerical computations.

\section{Definitions}
\label{sec:Definitions}

In this section, we first give the definition of the
symmetry-resolved entanglement entropy and mutual information for 
a subsystem composed of two disjoint regions. We explain their
relation with the multi-charged moments of the reduced density matrix, 
and we introduce the replica method to calculate them in CFTs.

\subsection{Symmetry-resolved entanglement entropies and mutual information}\label{sub:symi}
 
As we already pointed out in Sec.~\ref{sec:intro}, we take a spatial bipartition $A\cup B$ of an extended quantum system in a pure
state $|\Psi\rangle$, with $A$ made of two disconnected regions, $A=A_1\cup A_2$. We assume that the system is endowed
with a global $U(1)$ symmetry generated by a local charge $Q$. 
Given the partition of the system in different subsets, we can consider the charge operator in each of them; for example, in region $A$, it can be
obtained as $Q_A=\Tr_B(Q)$. If $|\Psi\rangle$ is an eigenstate 
of $Q$, the density matrix $\rho=\ket{\Psi}\bra{\Psi}$ commutes with $Q$, i.e. $[Q,\rho]=0$, and, by taking the trace over 
$B$, we find that $[Q_A,\rho_A]=0$. This implies that the reduced 
density matrix $\rho_A$ presents a block diagonal structure, in which each block 
corresponds to an eigenvalue $q\in \mathbb{Z}$ of $Q_A$. That is,
\begin{equation}
\label{eq:blockdecomposition}
\rho_A=\bigoplus_q \Pi_q \rho_A=\bigoplus_q \left[ p(q) \rho_A(q)\right],
\end{equation}
where $\Pi_q$ is the projector onto the eigenspace associated to the eigenvalue 
$q$ and $p(q)=\Tr\left(\Pi_q \rho_A\right)$ is the probability of obtaining $q$ as the outcome of a measurement of $Q_A$. Notice that Eq.~\eqref{eq:blockdecomposition} 
guarantees the normalisation $\mathrm{Tr}[\rho_A(q)]=1$ for any $q$.

The amount of entanglement between $A$ and $B$ in each symmetry sector can be 
quantified by the {\it symmetry-resolved R\'enyi entropies}, defined as
\begin{equation}
\label{eq:SRRE}
S_n^A(q)\equiv \frac{1}{1-n}\ln \mathrm{Tr}[\rho_A(q)^n].
\end{equation}
Taking the limit $n\to 1$ in this expression, we obtain the
{\it symmetry-resolved entanglement entropy}, 
\begin{equation}
\label{eq:SREE}
S_1^A(q)\equiv -\mathrm{Tr}[\rho_A(q)\ln \rho_A(q)].
\end{equation}
According to the decomposition of Eq.~\eqref{eq:blockdecomposition}, the total 
entanglement entropy in Eq.~(\ref{eq:defEE}) can be written as \cite{nc-10}
\begin{equation}
\label{eq:Ent_symm_decomp}
S_1^A=\sum_q p(q)S_1^A(q)-\sum_q p(q)\ln p(q)\equiv S_{\textrm{c}}^A+S_{\textrm{num}}^A,
\end{equation}
where $S_{\textrm{c}}$ is known as {\it configurational entropy} and quantifies the average contribution to the total entanglement of all the charge sectors \cite{fis,wv-03,bhd-18,bcd-19}, while 
$S_{\textrm{num}}$ is called {\it number entropy} and takes into account the entanglement due to the fluctuations of the value of the charge within the subsystem $A$ 
\cite{fis,kusf-20,kusf-20b,ms-20,kufs-21,zshgs-20,kufs-21b}.

Since the total charge in $A$ is the sum of the charge in $A_1$ and $A_2$, 
$Q_A=Q_{A_1}+Q_{A_2}$, then the reduced density matrices $\rho_{A_1}$, 
$\rho_{A_2}$ of $A_1$ and $A_2$ can be independently decomposed in charged 
sectors as we did for $\rho_A$ in Eq.~\eqref{eq:blockdecomposition}. 
Therefore, we can define the symmetry-resolved entropies $S_n^{A_1}(q_1)$, $S_n^{A_2}(q_2)$ for the regions 
$A_1$ and $A_2$ analogous to Eq.~\eqref{eq:SRRE} for $A$, with $q=q_1+q_2$. In Ref.~\cite{pbc-21}, it 
has been proposed to define the {\it symmetry-resolved mutual information} as
\begin{equation}
\label{eq:SRMI}
I^{A_1: A_2}(q)=\sum_{q_1=0}^q p(q_1,q-q_1)\left[S_1^{A_1}(q_1)+S_1^{A_2}(q-q_1)\right]-S_1^{A}(q).
\end{equation}
The quantity $p(q_1,q-q_1)$, normalised as
\begin{equation}
\label{eq:norm_condprob}
\sum_{q_1=0}^q p(q_1,q-q_1)=1,
\end{equation}
is the probability that a simultaneous measurement of the charges $Q_{A_1}$ and 
$Q_{A_2}$ yields $q_1$ and $q-q_1$, respectively, while the charge of the 
whole system $A$ is fixed to $q$. Although Eq.~\eqref{eq:SRMI} is a natural 
definition, $I^{A_1: A_2}(q)$ is not in general a good measure of the total correlations
between $A_1$ and $A_2$ within each charge sector since, in some cases, it can be 
negative \cite{pbc-21}. Nevertheless, we find interesting to investigate 
this quantity given that it provides a decomposition for the total mutual information 
(\ref{eq:SRMI}) similar to the one reported in Eq.~\eqref{eq:Ent_symm_decomp} for the 
entanglement entropy,
 \begin{equation}
\label{eq:totalMI}
I^{A_1: A_2}=\sum_q p(q) I^{A_1: A_2}(q) +
I^{A_1: A_2}_{\textrm{num}},
\end{equation}
where $I^{A_1: A_2}_{\textrm{num}}\equiv S^{A_1}_{\textrm{num}}
+S^{A_2}_{\textrm{num}}-S^{A}_{\textrm{num}} $ is the {\it number mutual information}.

\subsection{Charged moments and symmetry resolution}

The computation of the symmetry-resolved entanglement entropies and
mutual information from the definitions (\ref{eq:SRRE}) 
and (\ref{eq:SRMI}) requires the knowledge of the entanglement spectra 
of $\rho_A$, $\rho_{A_1}$ and $\rho_{A_2}$ and their symmetry resolution. 
However, this is usually a very difficult task, in particular if one is interested 
in analytical expressions. Alternatively, one can employ the {\it charged moments}
of the reduced density matrices. For $\rho_A$, they are defined as
\begin{equation}\label{eq:Charged_alpha}
 Z_n^{A}(\alpha)=\Tr[\rho_A^n e^{i\alpha Q_A}].
\end{equation}
Similar quantities can also be introduced for the two subsystems
$A_1$ and $A_2$ that constitute $A$ by replacing $\rho_A$ and $Q_A$ by
$\rho_{A_p}$ and $Q_{A_p}$, $p=1,2$. If we take now their Fourier transform,
\begin{equation}
\label{eq:Znq_alpha}
\mathcal{Z}_n^{A}(q)=\int_{-\pi}^\pi\frac{{\rm d} \alpha }{2\pi}e^{-i\alpha q} Z_n^{A}(\alpha),
\end{equation}
the symmetry-resolved entanglement entropies of the subsystem $A$ are given by~\cite{goldstein, xavier}
\begin{equation}
\label{eq:SRREfromqmoments}
S_n^{A}(q)=\frac{1}{1-n}\ln \left[
\frac{\mathcal{Z}_n^{A}(q)}{\left(\mathcal{Z}_1^{A}(q)\right)^n}
\right].
\end{equation}
In a similar manner, replacing $A$ with $A_1$ and $A_2$, 
we can obtain the symmetry-resolved entanglement entropies of the 
two components of $A$.

Notice that, for computing the symmetry-resolved mutual information of Eq.~\eqref{eq:SRMI}, we 
need to determine $p(q_1, q-q_1)$, i.e. the probability 
that a measurement of $Q_{A_1}$ and $Q_{A_2}$ gives $q_1$ and $q-q_1$ 
respectively, with $Q_A$ fixed to $q$. In order to calculate 
it, we consider the generalisation of the charged moments in Eq.~\eqref{eq:Charged_alpha} 
introduced for the first time in Ref.~\cite{pbc-21},
\begin{equation}
\label{eq:Charged_alphabeta}
Z_n^{A_1:A_2}(\alpha,\beta)=\mathrm{Tr}\left[\rho_A^n e^{i\alpha Q_{A_1}+i\beta Q_{A_2}}\right].
\end{equation}
We refer to them as {\it multi-charged moments}.
When $\alpha=\beta$, Eq.~\eqref{eq:Charged_alphabeta} reduces to the
charged moments of $A=A_1\cup A_2$ of Eq.~\eqref{eq:Charged_alpha}. 
If we take the Fourier transform of Eq.~(\ref{eq:Charged_alphabeta}), 
\begin{equation}
\label{eq:Znq_alphabeta}
\mathcal{Z}_n^{A_1:A_2}(q_1,q_2)=\int_{-\pi}^\pi\frac{{\rm d} \alpha }{2\pi}\frac{{\rm d} \beta }{2\pi} e^{-i\alpha q_1-i\beta q_2}
Z_n^{A_1:A_2}(\alpha,\beta),
\end{equation}
then $\mathcal{Z}_1^{A_1:A_2}(q_1, q_2)$ can be interpreted as the probability of having $q_1$ and $q_2$ as outcomes of a 
measurement of $Q_{A_1}$ and $Q_{A_2}$ respectively, independently of the value of 
$Q_A$. Therefore, it satisfies the normalisation
\begin{equation}
\label{eq:normalisationq1q2}
\sum_{q_1,q_2} \mathcal{Z}_1^{A_1:A_2}(q_1,q_2)=1,
\end{equation}
and $p(q_1, q-q_1)$ can be calculated as the conditional probability
\begin{equation}
\label{eq:CondProbability}
p(q_1,q-q_1)=\frac{\mathcal{Z}_1^{A_1: A_2}(q_1,q-q_1)}{p(q)},
\end{equation}
which fulfills Eq.~(\ref{eq:norm_condprob}).

\subsection{Charged moments in CFT}

In the rest of the paper, we will analyse the previous quantities in $1+1$-dimensional CFTs with a global
$U(1)$ symmetry. 
We will assume that the entire system is in the ground state and that 
the spatial dimension is an infinite line which we will divide 
into two parts $A$ and $B$, with $A$ made up of two disjoint intervals, 
namely $A=A_1\cup A_2=[u_1,v_1]\cup [u_2,v_2]$. If we denote by $\ell_1$ 
and $\ell_2$ the lengths of the two intervals and $d$ their separation, we have
\begin{equation}
\label{eq:elld_def}
\ell_1=|v_1-u_1|,\qquad \ell_2=|v_2-u_2|,\qquad d=|u_2-v_1|,\qquad x=\frac{\ell_1\ell_2}{(d+\ell_1)(d+\ell_2)},
\end{equation}
where we have also introduced the cross ratio $x$ of the four end-points, which takes values between $0$ and $1$.

\begin{figure}[t!]
\centering
\includegraphics[width=0.48\textwidth]{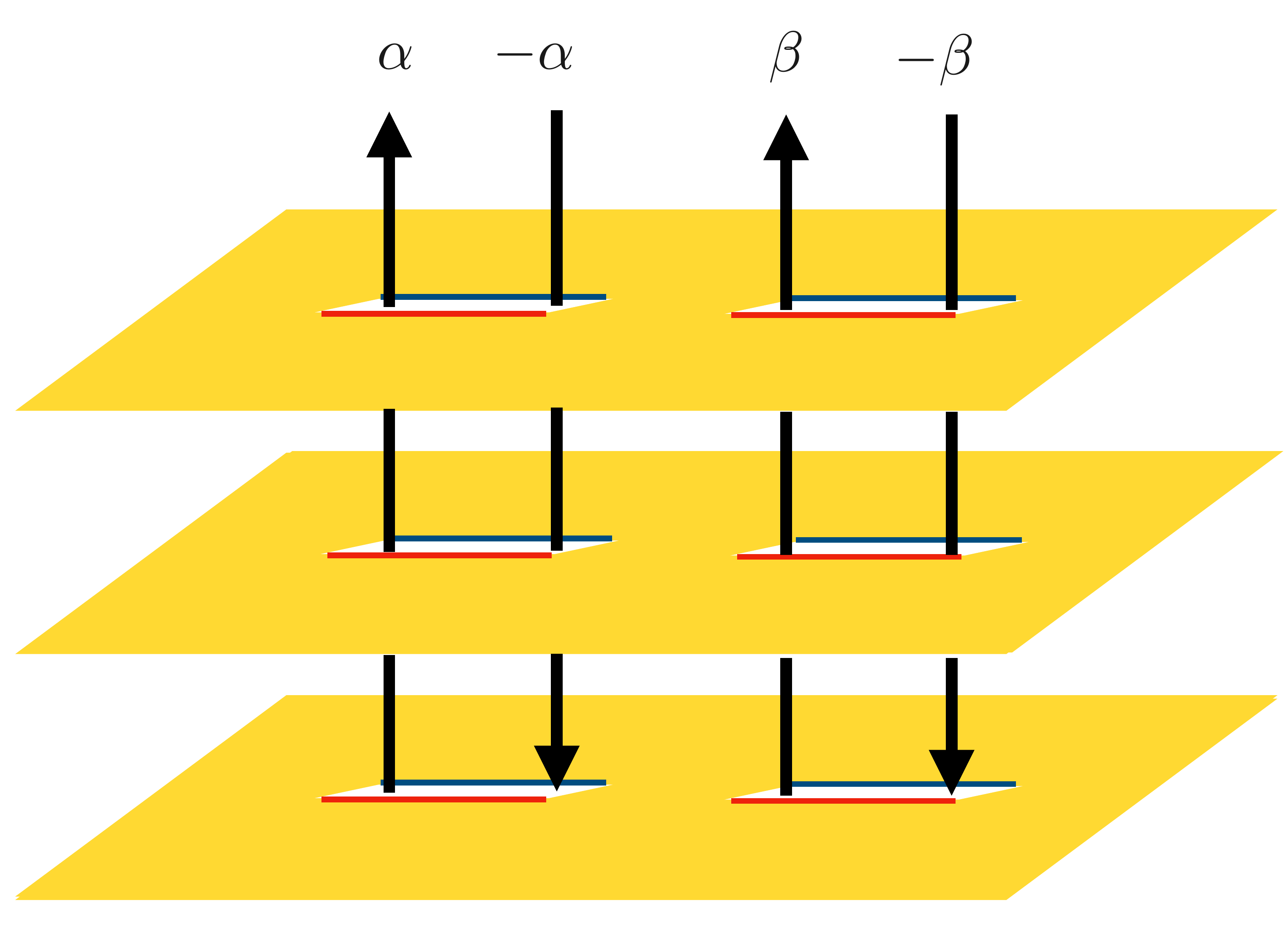}
\caption{Representation of the $n$-sheeted Riemann 
surface $\Sigma_n$ for $n=3$. The red edge of each cut is identified with the blue edge of the corresponding cut in the lower copy. The calculation of the multi-charged moments in Eq. \eqref{eq:Charged_alphabeta}
requires to insert different magnetic fluxes between the sheets, which we indicate by the arrows. The operator $e^{i\alpha Q_{A_1}}$ is implemented by the flux insertions $\alpha$ and $-\alpha$ along the left interval, while $e^{i\beta Q_{A_2}}$ corresponds to the fluxes $\beta$ and $-\beta$ at the right interval. }
\label{fig:r2}
\end{figure}

As explained in detail in Refs.~\cite{cc-04, cc-09}, using the path integral representation of $\rho_A$,
the moments $Z_n^{A}(0)=\Tr[\rho_A^n]$ are equal to the partition function of 
the CFT on a Riemann surface, which we call $\Sigma_n$, obtained as follows. We take the 
complex plane where the CFT is originally defined and we perform two cuts 
along the intervals $A_1=(u_1, v_1)$ and $A_2=(u_2, v_2)$. Then we replicate 
$n$ times the cut plane and we glue the copies together along the cuts in a cyclical way as we illustrate in Fig.~\ref{fig:r2}. We eventually obtain an $n$-sheeted 
Riemann surface of genus $n-1$, which is symmetric under the $\mathbb{Z}_n$ cyclic permutation of the
sheets. 

Alternatively, instead of replicating the space-time where the CFT is 
initially defined, one can take $n$ copies of the CFT on the complex plane and 
quotient it by the $\mathbb{Z}_n$ symmetry under the cyclic exchange of the copies. We then 
get the orbifold theory ${\rm CFT}^{\otimes n}/\mathbb{Z}_n$. The moments $Z_n^{A}(0)$
are equal to the four-point correlation function on the complex plane~\cite{cc-04,ccd-08},
\begin{equation}
 Z_n^A(0)=\langle \tau_n(u_1)\tilde{\tau}_n(v_1)\tau_n(u_2)\tilde{\tau}_n(v_2)\rangle,
\end{equation}
where $\tau_n$ and $\tilde{\tau}_n$ are dubbed as twist and anti-twist fields~\cite{k-87, dixon, cc-04,ccd-08}. They implement 
in the orbifold the multivaluedness of the correlation functions on the surface $\Sigma_n$
when we go around its branch points. In fact, the winding around the point where 
$\tau_n$ ($\tilde{\tau}_n$) is inserted maps a field $\mathcal{O}_k$ living
in the copy $k$ of the orbifold into the copy $k+1$ ($k-1$), that is
\begin{equation}
\tau_n(u) \mathcal{O}_k(e^{2\pi i}(z-u))= \mathcal{O}_{k+1}(z-u)\tau_n(u).
\end{equation}
The twist and anti-twist fields are spinless primaries with conformal weight
\begin{equation}
 h_n^{\tau}=\frac{c}{24}\left(n-\frac{1}{n}\right),
\end{equation}
where $c$ is the central charge of the initial CFT.

The charged moments of $\rho_A$ can also be computed employing the previous 
frameworks. As argued in Ref.~\cite{goldstein}, the operator $e^{i\alpha Q_A}$ can be 
interpreted as a magnetic flux between the sheets of the surface $\Sigma_n$, 
such that a charged particle moving along a closed path  that crosses all the 
sheets acquires a phase $e^{i\alpha}$.  For the multi-charged moments introduced in Eq.~\eqref{eq:Charged_alphabeta}, we have to insert two different magnetic fluxes 
$\alpha$ and $\beta$ at the cuts $A_1$ and $A_2$ respectively, as we pictorially show in Fig.~\ref{fig:r2}. They can be implemented 
by a local $U(1)$ operator $\mathcal{V}_\alpha(x)$ that generates a phase shift $e^{i\alpha}$
along the real interval $[x, \infty)$. Then the charged moments are equal to
the four-point correlation function on the surface $\Sigma_n$
\begin{equation}\label{eq:gen_charged_mom_corr_riemann}
 Z_n^{A_1:A_2}(\alpha,\beta)=Z_n^{A}(0)\langle \mathcal{V}_\alpha(u_1)\mathcal{V}_{-\alpha}(v_1)
 \mathcal{V}_\beta(u_2)\mathcal{V}_{-\beta}(v_2)\rangle_{\Sigma_n}. 
\end{equation}

In the orbifold theory, the magnetic flux can be incorporated by considering the 
composite twist field $\tau_{n,\alpha}\equiv\tau_n\cdot \mathcal{V}_\alpha$. Thus, 
if we take a field $\mathcal{O}_k$ in the copy $k$ of the orbifold, then the winding
$(z-u)\mapsto e^{2\pi i}(z-u)$ around the point $u$ where $\tau_{n,\alpha}$
is inserted gives rise to a phase $e^{i\alpha/n}$,
\begin{equation}\label{eq:composite_twist_field}
\tau_{n, \alpha}(u) \mathcal{O}_k(e^{2\pi i}(z-u))
 = e^{i\alpha/n}\mathcal{O}_{k+1}(z-u)\tau_{n, \alpha}(u).
\end{equation}
The same applies to the composite anti-twist field $\tilde{\tau}_{n,\alpha}\equiv \tilde{\tau}_n\cdot \mathcal{V}_\alpha$, which takes a field from the copy $k$ to $k-1$
adding a phase $e^{i\alpha/n}$.
Therefore, Eq.~\eqref{eq:gen_charged_mom_corr_riemann} can be re-expressed as the four-point function 
on the complex plane
\begin{equation}\label{eq:charged_mom_twist_corr}
 Z_n^{A_1:A_2}(\alpha,\beta)=\langle \tau_{n,\alpha}(u_1)\tilde{\tau}_{n,-\alpha}(v_1)
 \tau_{n,\beta}(u_2)\tilde{\tau}_{n,-\beta}(v_2)\rangle.
\end{equation}
In Ref.~\cite{goldstein}, it is shown that, if $\mathcal{V}_\alpha$ is a spinless primary
operator with conformal weight $h_\alpha^{\mathcal{V}}$, then so are the composite twist and anti-twist 
fields, with conformal weights
\begin{equation}
h_{n,\alpha}=h_n^{\tau}+\frac{h_\alpha^{\mathcal{V}}}{n}.
\end{equation}

One can further consider other configurations for the magnetic
fluxes between  the sheets of the Riemann surface $\Sigma_n$. In
general, if we assume that a particle gets a different phase 
$e^{i\alpha_j}$ when it goes around each branch point, provided they satisfy the neutrality condition 
$\alpha_1+\alpha_2+\alpha_3+\alpha_4=0$, then the partition function
of this theory is given by
\begin{equation}\label{eq:gen_partition_func}
 Z_n^{A_1:A_2}(\{\alpha_j\})=Z_n^{A}(0)\langle \mathcal{V}_{\alpha_1}(u_1)
 \mathcal{V}_{\alpha_2}(v_1)\mathcal{V}_{\alpha_3}(u_2)\mathcal{V}_{\alpha_4}(v_2)\rangle_{\Sigma_n},
\end{equation}
or, in terms of the composite twist fields, by
\begin{equation}\label{eq:twist_gen_partition_func}
 Z_n^{A_1:A_2}(\{\alpha_j\})=\langle \tau_{n,\alpha_1}(u_1)\tilde{\tau}_{n,\alpha_2}(v_1)
 \tau_{n,\alpha_3}(u_2)\tilde{\tau}_{n,\alpha_4}(v_2)\rangle.
\end{equation}
Then the multi-charged moments $Z_n^{A_1:A_2}(\alpha,\beta)$ can
be treated as the particular case in which $\alpha_1=-\alpha_2=\alpha$ and, due to the neutrality condition, $\alpha_3=-\alpha_4=\beta$.

In Sec.~\ref{sec:FreeDirac}, we will compute the charged moments $Z_n^{A_1:A_2}(\alpha,\beta)$---and more in general the 
partition functions $Z_n^{A_1:A_2}(\{\alpha_j\})$---for 
the massless Dirac fermion using the orbifold theory ${\rm CFT}^{\otimes n}/\mathbb{Z}_n$. On the other hand, 
in Sec.~\ref{sec:CompactBoson}, we will adopt a geometric approach to obtain the multi-charged moments of the compact boson from
the correlation function on the Riemann surface
$\Sigma_n$ of Eq.~\eqref{eq:gen_partition_func}.

\section{Free massless Dirac field theory}
\label{sec:FreeDirac}

The massless Dirac field theory is described by the action
\begin{equation}
\label{eq:lagrangianDirac}
\mathcal{S}_{\rm D}=\int {\rm d}x_0 {\rm d}x_1 \bar{\psi}\gamma^\mu \partial_\mu \psi ,
\end{equation}  
where $\bar{\psi}=\psi^\dagger \gamma^0$. The $\gamma^\mu$ matrices can be represented 
in terms of the Pauli matrices as $\gamma^0=\sigma_1$ and $\gamma^1=\sigma_2$.
The action of Eq.~(\ref{eq:lagrangianDirac}) exhibits a global $U(1)$ symmetry: it is invariant 
if the fields are multiplied by a phase, i.e. $\psi\mapsto e^{i \alpha}\psi$ 
and $\bar{\psi}\mapsto e^{-i \alpha}\bar{\psi}$. By Noether's theorem, this symmetry 
is related to the conservation of the charge $Q_{\rm D}=\int {\rm d}x_1 \psi^\dagger\psi $.

 The ground state entanglement of a subsystem $A$ made up of multiple
 disjoint intervals in the ground state of this theory was first investigated in Ref.~\cite{CFH}. For the case of two disjoint 
 intervals, $A=A_1\cup A_2$, it was found that the moments of $\rho_A$ are
 \begin{equation}
 \label{eq:Dirac2intCFH}
 Z_n^A(0)
 =
 c_{n}\left[\ell_1 \ell_2 (1-x)\right]^{\frac{1-n^2}{6n}}
 ,
 \end{equation}
where $c_{n}$ is a non-universal constant.

In this section, we will compute the multi-charged moments of Eq.~(\ref{eq:Charged_alphabeta}) in the ground state of the massless
Dirac field theory.
We will extend the approach introduced in 
Ref.~\cite{CFH} for the moments of Eq.~\eqref{eq:Dirac2intCFH}. Similar techniques have been exploited 
in Ref.~\cite{MDC-20} for studying the charged moments of Eq.~(\ref{eq:Charged_alpha}), when
$A$ is a single interval, in two dimensional free massless Dirac theories and in Ref.~\cite{MBC-21} in the context of 
the charge imbalance resolved negativity. We will benchmark our analytical results with exact numerical 
calculations in a lattice model.
\subsection{Charged moments} 
In Sec.~\ref{sec:Definitions}, we explained that the partition function $Z_n^{A_1:A_2}(\{\alpha_j\} )$
can be obtained either by 
considering the theory on a complicated Riemann surface or by replicating  it $n$-times
and working with the orbifold on the complex plane. For the massless Dirac field theory, 
 the latter approach is more convenient. Thus, let us take the $n$-component field
\begin{equation}
\label{eq:nfieldsonC}
\Psi=
\begin{pmatrix}
\psi_1
\\
\psi_2
\\
\vdots 
\\
\psi_n
\end{pmatrix},
\end{equation}
where $\psi_j$ is the Dirac field on the $j$-th copy of the system.
Eq.~\eqref{eq:composite_twist_field} describes the effect of the composite twist 
fields on the components of $\Psi$ when going around the end-points of the subsystem
$A=A_1\cup A_2$. This transformation can be encoded in the
matrix
\begin{equation}
\label{eq:twistmatrix}
T_{a}=
\begin{pmatrix}
0 & e^{i a/n} & &
\\
 & 0 & e^{i a/n} &
\\
 & & \ddots & \ddots
 \\
 (-1)^{n-1}e^{i a/n} & & 0
\end{pmatrix}.
\end{equation}
In the general case of Eq.~\eqref{eq:twist_gen_partition_func},
$\Psi$ transforms according to $T_{\alpha_{2p-1}}$ when winding around
the point $u_p$ and to the transpose matrix $T_{\alpha_{2p}}^t$ when 
going around the point $v_p$, with $p=1, 2$.
The matrix $T_a$ in Eq.~(\ref{eq:twistmatrix}), sometimes called {\it twist matrix}, 
was  introduced for the case $a=0$ in Refs.~\cite{CFH,ccd-08} and for general $a$ in Ref. \cite{MDC-20}. Its eigenvalues 
are of the form
\begin{equation}
t_k=e^{i a/n} e^{2\pi i k/n},
\qquad
k=-\frac{n-1}{2}, \dots,\frac{n-1}{2}.
\end{equation}
By simultaneously diagonalising all the $T_{\alpha_j}$ with a unitary transformation (which is independent of $\alpha_j$), 
we can recast the replicated theory in $n$ decoupled 
fields $\psi_k$ on the plane, which are multi-valued,
\begin{eqnarray}\label{eq:Dirac_k_bc}
 \psi_k(e^{i2\pi}(z-u_p))&=&e^{i\alpha_{2p-1}/n}e^{2\pi i k/n}\psi_k(z-u_p),\nonumber \\ 
 \psi_k(e^{i2\pi}(z-v_p))&=&e^{i\alpha_{2p}/n}e^{-2\pi i k/n}\psi_k(z-v_p).
\end{eqnarray}
Notice that this technique, known as {\it diagonalisation in the replica space}, can be applied 
only to free theories since, otherwise, the $k$-modes do not decouple. For the free massless 
Dirac theory, this allows us to write the Lagrangian of the replicated theory as
\begin{equation}
\mathcal{L}_{{\rm D}, n}=\sum_k \mathcal{L}_k,
\qquad
\mathcal{L}_k= \bar{\psi}_k\gamma^\mu \partial_\mu \psi_k.
\end{equation}
 Following this approach, the partition function of Eq.~(\ref{eq:twist_gen_partition_func}) factorises into
\begin{equation}
\label{eq:ChargedMom4fluxestotal}
Z_{n}^{A_1:A_2}(\{\alpha_j\})=
\prod_{k=-\frac{n-1}{2}}^{k=\frac{n-1}{2}}Z_{k,n}^{A_1:A_2}(\{\alpha_j\}),
\end{equation}
where $Z_{k,n}^{A_1:A_2}(\{\alpha_j\})$ is the partition function for a Dirac field $\psi_k$ with
the boundary conditions of Eq.~\eqref{eq:Dirac_k_bc}.

The main difference between the partition functions $Z_{k,n}^{A_1:A_2}(\{\alpha_j\})$ and the standard computation of Ref.~\cite{CFH} 
for R\'enyi entropies is that 
the boundary conditions of the multi-valued fields around the branch points now depend on the phases $\alpha_j$.
This multivaluedness can be removed, as done in \cite{CFH} for $\alpha_j=0$, by introducing an 
external $U(1)$ gauge field $A_\mu^k$ coupled to single-valued fields $\tilde{\psi}_k$.
In fact, if we apply the singular gauge transformation 
\begin{equation}
\label{eq:singulargaugetrans}
\psi_k(x)=e^{i\int_{x_0}^{x}\mathrm{d}y^\mu A_\mu^k} \tilde{\psi}_k(x),
\end{equation}
then the Lagrangian for the $k$-th mode can be rewritten as
\begin{equation}
\label{eq:Diracaction_singularpsi}
\mathcal{L}_k=\bar{\tilde{\psi}}_k\gamma^\mu\left( \partial_\mu +i A_\mu^k\right)\tilde{\psi}_k,
\end{equation}
with the advantage of absorbing the phase 
around the end-points of $A_1\cup A_2$ into the gauge field.
The only requirement that $A_\mu^k$ in Eq.~(\ref{eq:singulargaugetrans}) must satisfy
is that, integrated along any closed curve $\mathcal{C}$ that encircles the end-points 
of $A$, the boundary conditions of Eq.~\eqref{eq:Dirac_k_bc} for $\psi_k$ must be reproduced.  
For this purpose, we require
\begin{equation}
\label{eq:monodromy_int1}
\oint_{\mathcal{C}_{u_p}} {\rm d} y^\mu A_\mu^k= -\frac{2\pi k}{n}-\frac{\alpha_{2p-1}}{n} ,
\qquad
\oint_{\mathcal{C}_{v_p}} {\rm d} y^\mu A_\mu^k= \frac{2\pi k}{n}-\frac{\alpha_{2p}}{n},
\end{equation}
where $\mathcal{C}_{u_p}$ and $\mathcal{C}_{v_p}$ are closed contours around the end-points of the $p$-th interval. Moreover, we have to impose that, if $\mathcal{C}$ does not enclose any end-point, then $\oint_{\mathcal{C}} {\rm d} y^\mu A_\mu^k=0$. Applying the Stoke's theorem, the conditions of Eqs.~(\ref{eq:monodromy_int1}) 
can be expressed in differential form,
\begin{equation}
\label{eq:monodromy_diff}
\frac{1}{2\pi}\epsilon^{\mu \nu}\partial_\nu A_\mu^k(x)=
\sum_{p=1}^2
\left[
\left(\frac{\alpha_{2p-1}}{2\pi n}+\frac{k}{n}\right)\delta(x-u_p)+
\left(\frac{\alpha_{2p}}{2\pi n}-\frac{k}{n}\right)\delta(x-v_p)\right]
.
\end{equation}
Once the transformation of Eq.~(\ref{eq:singulargaugetrans}) is performed, the partition function $Z_{k,n}^{A_1:A_2}(\{\alpha_j\})$ of the $k$-th mode is equal to the vacuum expectation value
\begin{equation}
\label{eq:partition_gauge}
Z_{k,n}^{A_1:A_2}(\{\alpha_j\})=   \left\langle
e^{i\int d^2 x j^\mu_k A_\mu^k}
\right\rangle ,
\end{equation}
where $j^\mu_k\equiv \bar{\tilde{\psi}}_k\gamma^\mu\tilde{\psi}_k$ is the conserved Dirac current
for each mode.
Eq.~(\ref{eq:partition_gauge}) can be easily computed via bosonisation~\cite{CFH}, which allows us to
write the current in terms of the dual scalar field $\phi_k$ such that $j^\mu_k=\epsilon^{\mu\nu}\partial_\nu\phi_k/\sqrt{\pi}$.
If we use this result in Eq.~\eqref{eq:partition_gauge}, and we apply Eq.~\eqref{eq:monodromy_diff}, then 
$Z_{k,n}^{A_1:A_2}(\{\alpha_j\})$ is equal to the following correlation function of vertex operators $V_a(y)=e^{-i a\phi_k(y)}$
\begin{equation}
\label{eq:ChargedMom4fluxes}
Z_{k,n}^{A_1:A_2}(\{\alpha_j\})=\langle  V_{\frac{k}{n}+\frac{\alpha_1}{2\pi n}}(u_1) V_{-\frac{k}{n}+\frac{\alpha_2}{2\pi n}}(v_1) V_{\frac{k}{n}+\frac{\alpha_3}{2\pi n}}(u_2) V_{-\frac{k}{n}+\frac{\alpha_4}{2\pi n}}(v_2)\rangle.
\end{equation}
Notice that the neutrality condition  $\alpha_1+\alpha_2+\alpha_3+\alpha_4=0$ ensures that the latter  correlator does not vanish.
The correlation function of vertex operators in the complex plane is well-known 
(see, for instance, Ref.~\cite{difrancesco}) and, therefore,  Eq.~(\ref{eq:ChargedMom4fluxes}) 
can be easily calculated. Plugging the result into Eq.~(\ref{eq:ChargedMom4fluxestotal}) 
and performing the product over $k$, we obtain
\begin{multline}
\label{eq:ChargedMom4fluxestotal_v2}
Z_{n}^{A_1:A_2}(\{\alpha_j\})
\propto\\
\left[\ell_1 \ell_2 (1-x) \right]^{\frac{1-n^2}{6n}} 
\left[d^{\alpha_2 \alpha_3}
\ell_1^{\alpha_1 \alpha_2}
\ell_2^{\alpha_3 \alpha_4}
(d+\ell_1)^{\alpha_1 \alpha_3}
(d+\ell_2)^{\alpha_2 \alpha_4}
(d+\ell_1+\ell_2)^{\alpha_1 \alpha_4}\right]^{\frac{1}{2\pi^2 n}},
\end{multline}
where $x$ is the cross-ratio defined in Eq.~(\ref{eq:elld_def}).
When we take $\alpha_1=-\alpha_2=\alpha$ and $\alpha_3=-\alpha_4=\beta$ in this expression,  we get the multi-charged
moments in Eq.~(\ref{eq:Charged_alphabeta}) as
\begin{eqnarray}
\label{eq:Charged_alphabeta_disj}
Z_n^{A_1:A_2}(\alpha,\beta)&=&c_{n;\alpha,\beta}\left[\ell_1 \ell_2 (1-x)\right]^{\frac{1-n^2}{6n}} 
\left[\left(1-x\right)^{-\alpha \beta}
\ell_1^{-\alpha^2 }
\ell_2^{-\beta^2}
\right]^{\frac{1}{2\pi^2 n}}.
\end{eqnarray}
We assume that all the length scales in this formula have been 
regularised through a UV cutoff which is included in the multiplicative constant $c_{n;\alpha,\beta}$.
\begin{figure}[htbp!]
\centering
\subfigure
{\includegraphics[width=0.48\textwidth]{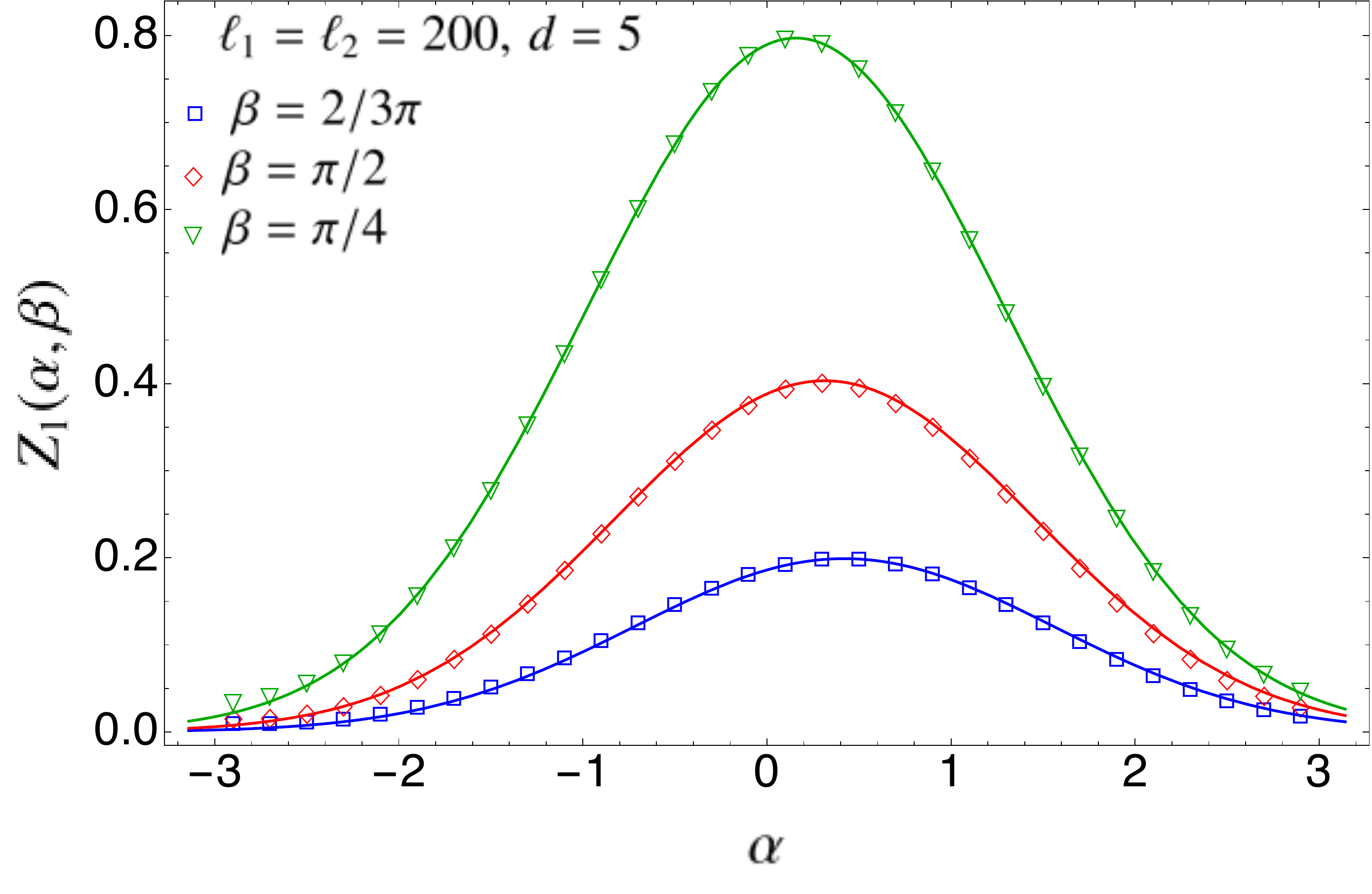}}
\subfigure
{\includegraphics[width=0.48\textwidth]{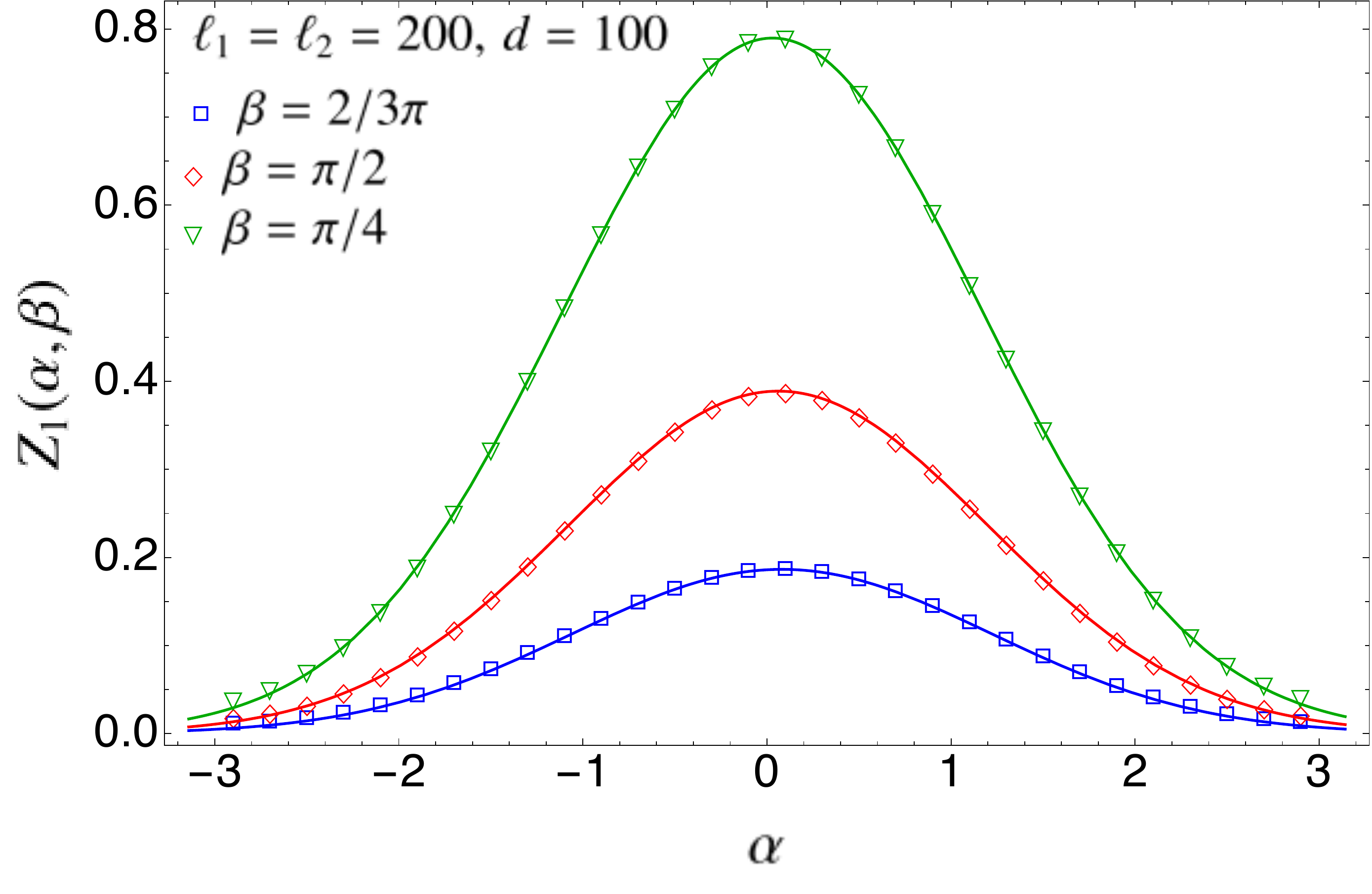}}
\subfigure
{\includegraphics[width=0.48\textwidth]{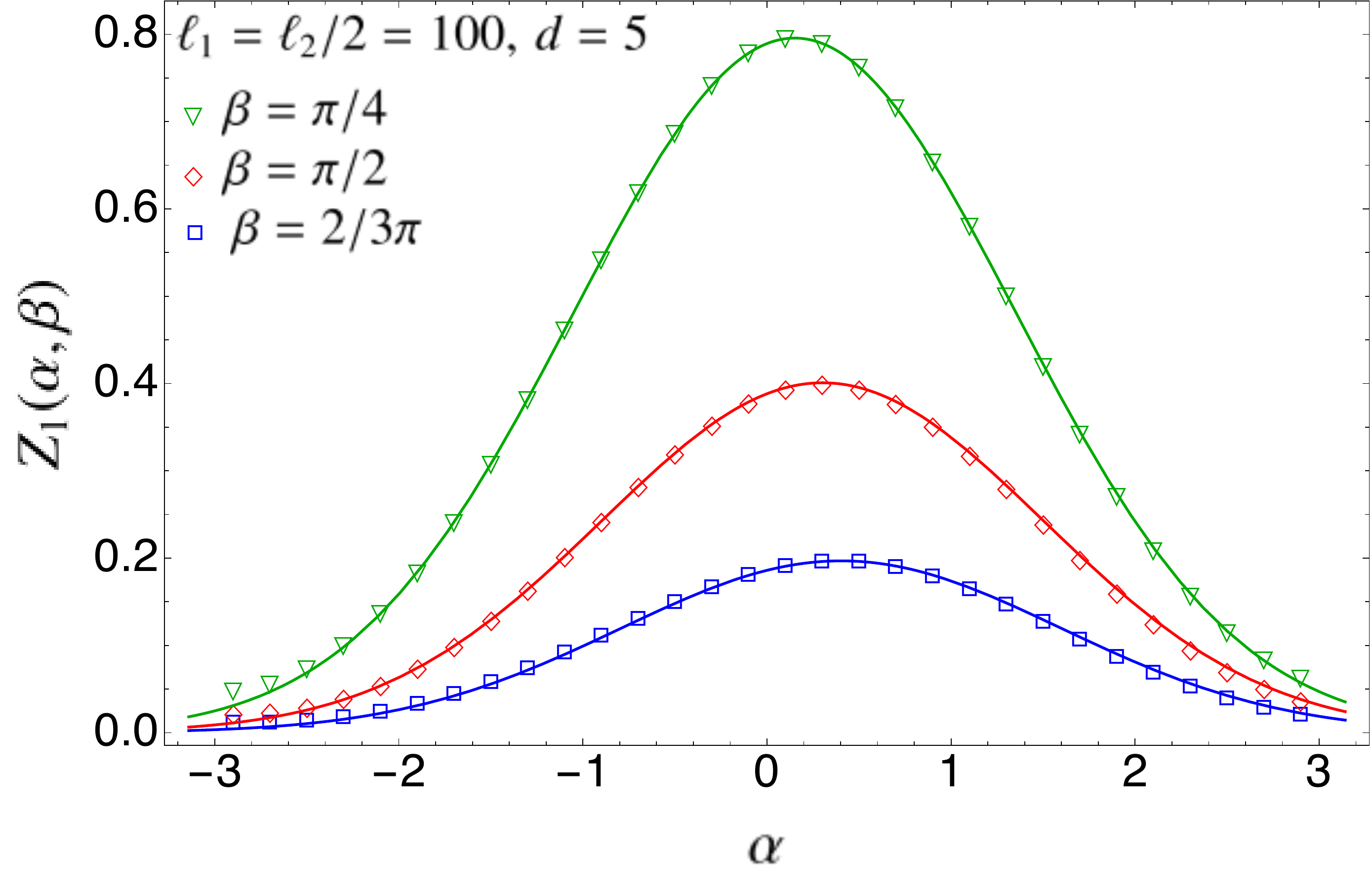}}
\subfigure
{\includegraphics[width=0.48\textwidth]{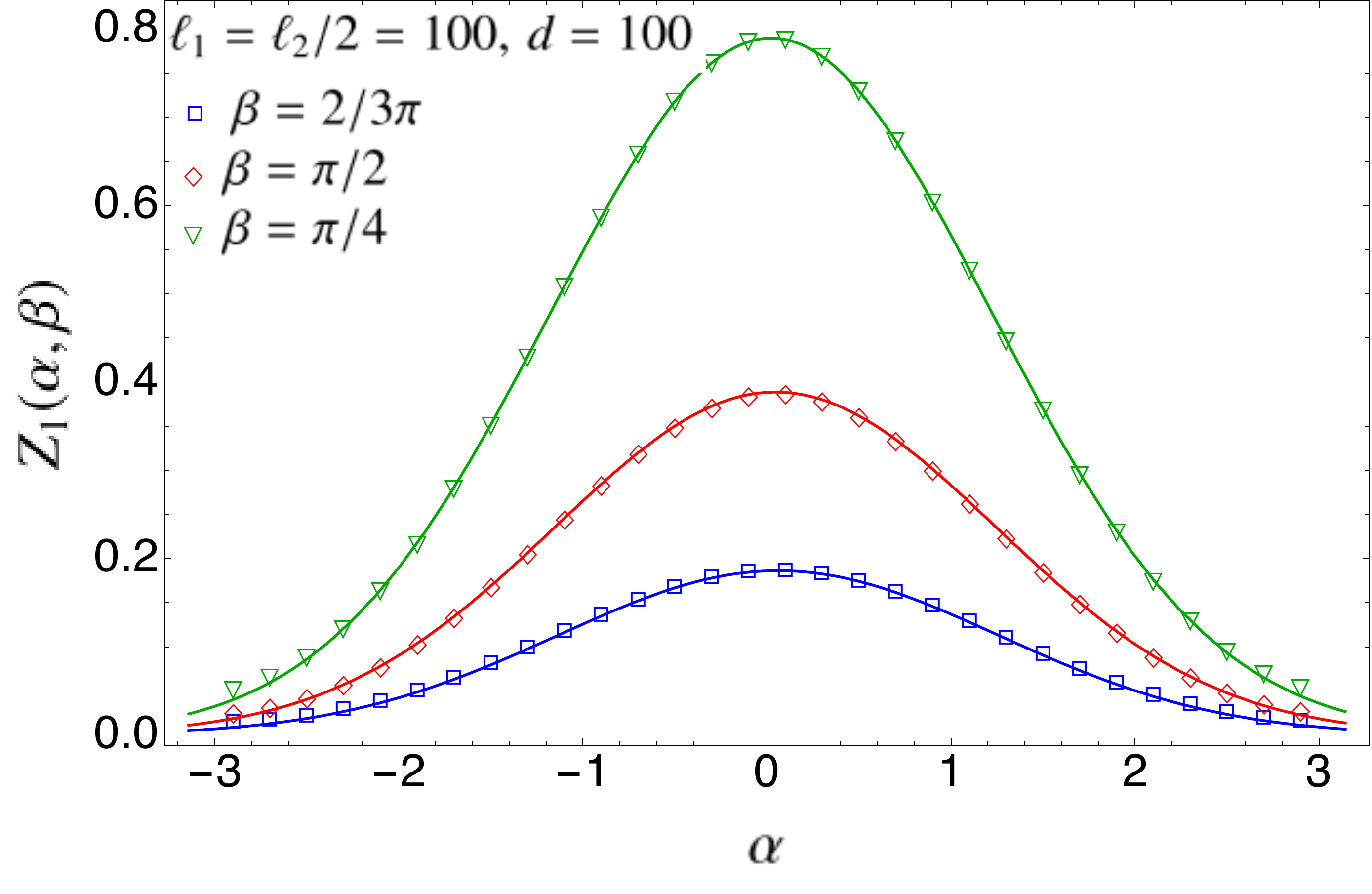}}
\subfigure
{\includegraphics[width=0.48\textwidth]{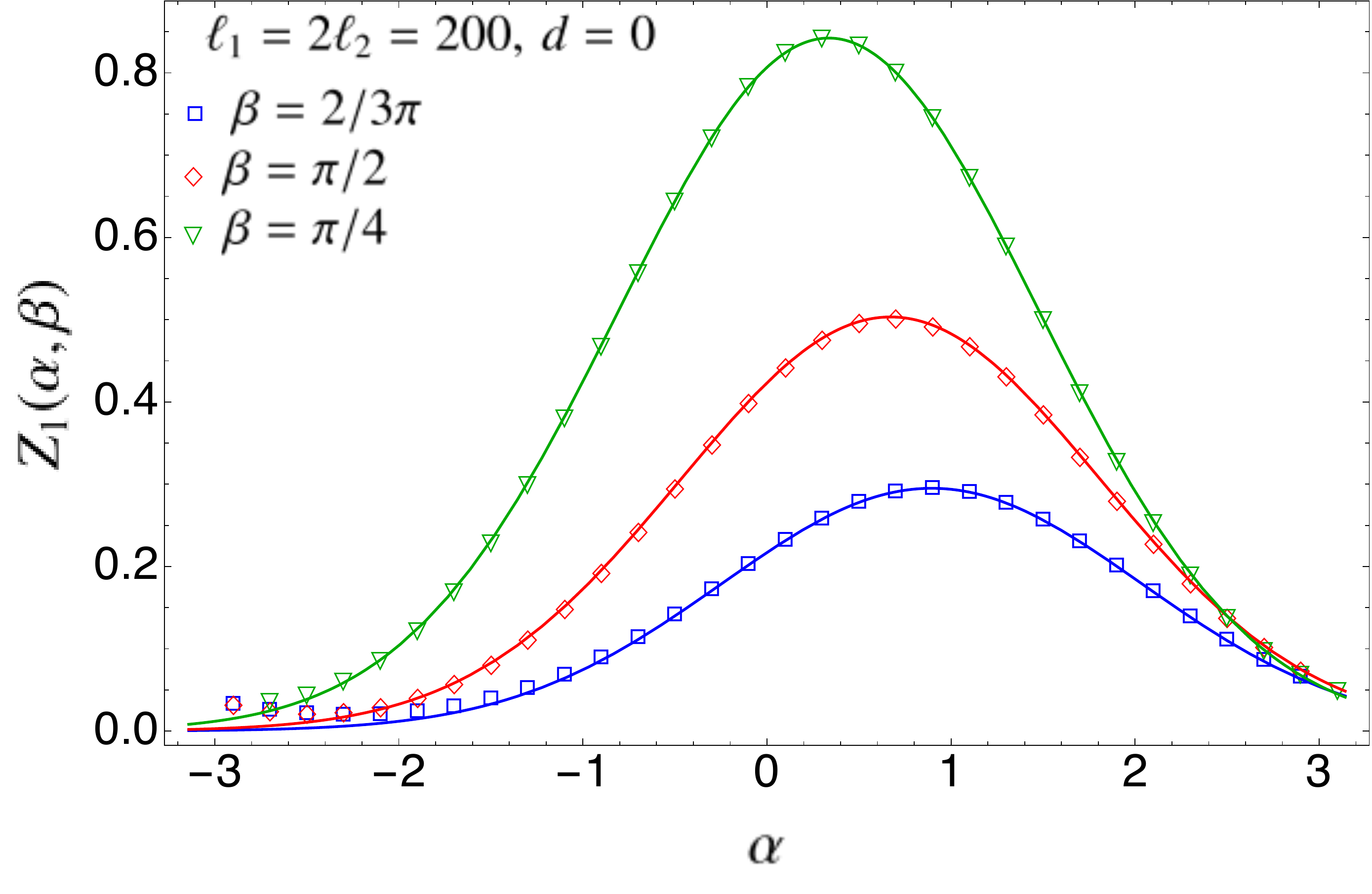}}
\subfigure
{\includegraphics[width=0.48\textwidth]{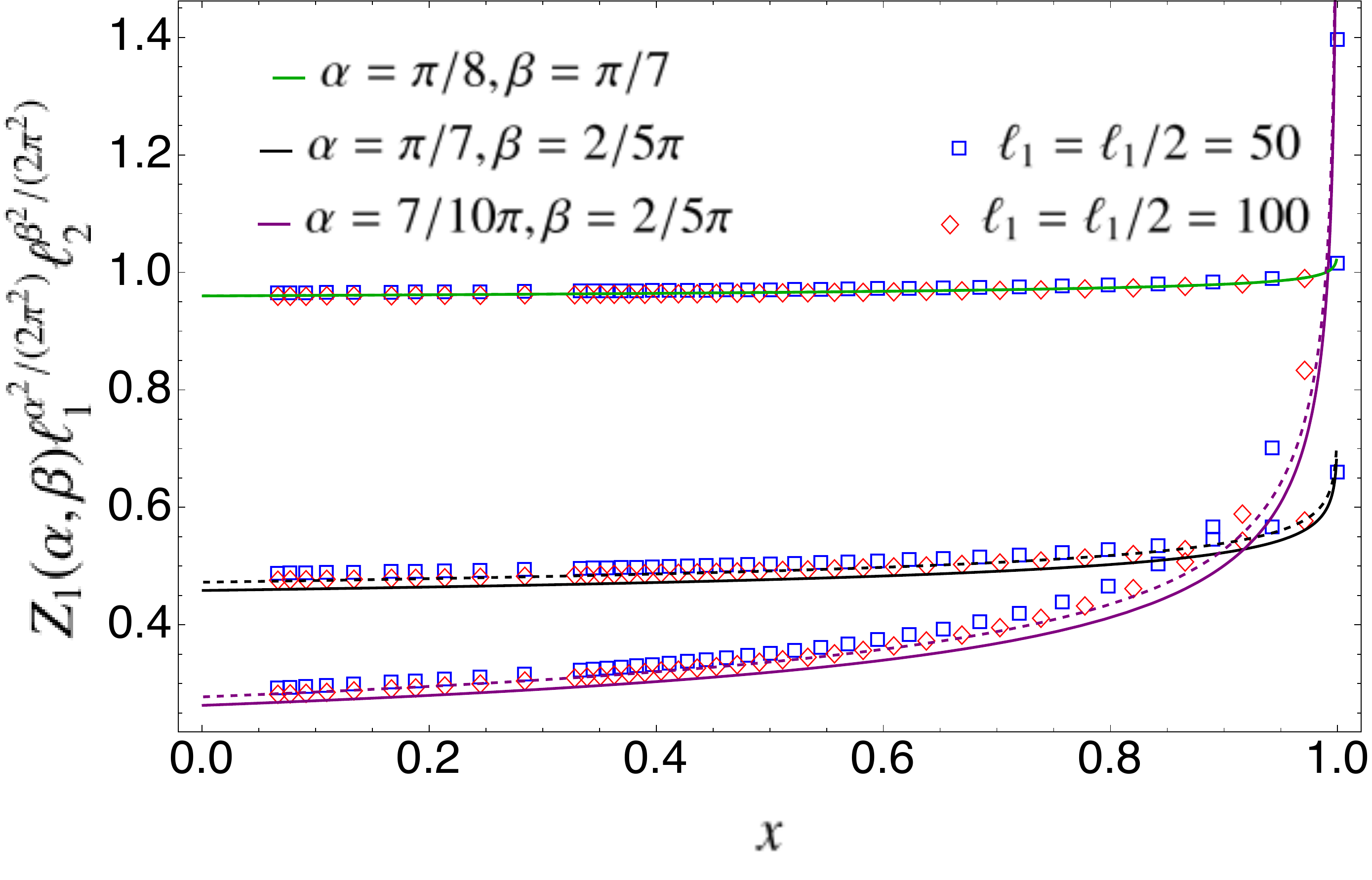}}
\caption{Analysis of $Z_1^{A_1:A_2}(\alpha,\beta)$ for the tight-binding model.
In the top and middle panels, we show $Z_1^{A_1:A_2}(\alpha,\beta)$ as a function of $\alpha$ at fixed $\beta$ for different intervals of lengths $\ell_1,\ell_2$, separated by a distance $d>0$. In this case, the solid lines are the theoretical predictions of Eq. \eqref{eq:Charged_alphabeta_disj}, taking Eq. \eqref{eq:FHalphabeta} for $c_{n;\alpha,\beta}$. In the bottom left panel, we repeat the analysis but considering adjacent intervals ($d=0$) and the solid curves correspond to Eq. \eqref{eq:Charged_alphabeta_adj}, with $\tilde{c}_{n;\alpha,\beta}$ conjectured in Eq. \eqref{eq:FHalphabeta_adj}. In the bottom right panel, we plot $Z_1^{A_1:A_2}(\alpha,\beta)$ as a function of the cross-ratio $x$. Here the solid lines correspond to Eq. \eqref{eq:Charged_alphabeta_disj} using the exact expression for $c_{n;\alpha,\beta}$ of Eq. \eqref{eq:FHalphabeta}, while for the dashed curves we have considered instead the quadratic approximation for this constant of Eq.~\eqref{eq:quadratic_non_univ_const}. In all
the cases, the points are the exact numerical values for $Z_1^{A_1:A_2}(\alpha,\beta)$ calculated as described in Appendix~\ref{app:numericaltools}.}
\label{fig:Chargedmomentsdnon0}
\end{figure}

An interesting case to analyse is when the two intervals $A_1$ and $A_2$ 
become adjacent; that is, when $d\to0$. In that limit, the cross-ratio $x$ tends to
one such that 
\begin{equation}\label{eq:cross_ratio_adj}
    1-x=d\frac{\ell_1+\ell_2}{\ell_1\ell_2}+O(d^2),
\end{equation}
and Eq.~\eqref{eq:Charged_alphabeta_disj} vanishes. Nevertheless, in this regime,
the distance $d$ must be regarded as another UV cutoff, which can be absorbed in
the multiplicative constant $c_{n;\alpha,\beta}$. Therefore, from Eqs.~\eqref{eq:cross_ratio_adj} and (\ref{eq:Charged_alphabeta_disj}), one expects
\begin{equation}
\label{eq:Charged_alphabeta_adj}
Z_n^{A_1:A_2}(\alpha,\beta)=
\tilde{c}_{n;\alpha,\beta}(\ell_1+\ell_2)^{\frac{1-n^2}{6n}}\left[\frac{\ell_1^{\alpha \beta-\alpha^2 }
\ell_2^{\alpha \beta-\beta^2}}{(\ell_1+\ell_2)^{\alpha \beta}}
\right]^{\frac{1}{2\pi^2 n}},
\end{equation}
which agrees with the fact that, according to Eq.~\eqref{eq:charged_mom_twist_corr}, the 
multi-charged moments $Z_n^{A_1:A_2}(\alpha,\beta)$ must tend to
the three-point function of primaries 
$\langle \tau_{n,\alpha}(u_1)\mathcal{V}_{-\alpha+\beta}(v_1)\tilde{\tau}_{n,-\beta}(v_2)\rangle$, in the limit $d\to0$.


In Fig.~\ref{fig:Chargedmomentsdnon0}, we check the expressions obtained in Eqs.~\eqref{eq:Charged_alphabeta_disj}~and ~\eqref{eq:Charged_alphabeta_adj} for $Z_n^{A_1:A_2}(\alpha,\beta)$ 
with exact numerical calculations performed in the tight-binding model, which is a chain of non-relativistic free fermions whose scaling limit is described by the massless Dirac field theory. The details of the numerical 
techniques employed are given in Appendix \ref{app:numericaltools}. In order to compare Eqs.~(\ref{eq:Charged_alphabeta_disj}) 
and (\ref{eq:Charged_alphabeta_adj}) with the numerical data, we need the 
concrete expression of the non-universal factors $c_{n;\alpha,\beta}$ and $\tilde{c}_{n;\alpha,\beta}$ for this particular model. 
When the two intervals $A_1$ and $A_2$ are far away, that is in the limit $d\to\infty$, the 
multi-charged moments of $A_1\cup A_2$ factorise into those of $A_1$ and $A_2$,
\begin{equation}\label{eq:far_intervals}
 \lim_{d\to \infty} Z_{n}^{A_1:A_2}(\alpha, \beta)= Z_{n}^{A_1}(\alpha)Z_{n}^{A_2}(\beta).
\end{equation}
Therefore, one expects $c_{n;\alpha,\beta}$ to be the product of the 
two non-universal constants associated to $A_1$ and $A_2$ as single intervals. The latter were 
obtained for the tight-binding model in Ref.~\cite{riccarda} by exploiting the asymptotic properties of Toeplitz determinants. 
We can also apply here those results, taking into account that each interval is associated to a different flux, either $\alpha$ or $\beta$. Then we have
 \begin{equation}
\label{eq:FHalphabeta}
c_{n;\alpha,\beta}= e^{\left[-\frac{1}{3}\left(n-\frac{1}{n}\right)-\frac{\alpha^2}{2\pi^2 n}-\frac{\beta^2}{2\pi^2 n}\right]\log 2+\Upsilon(n,\alpha)+\Upsilon(n,\beta)},
\end{equation}
where \cite{riccarda}
\begin{equation}
\Upsilon{(n,\alpha)}= {n i}\int_{-\infty}^\infty  {\rm d}w [\tanh( \pi w)-\tanh (\pi n w+i\alpha/2)]  \ln \frac{\Gamma(\frac12 +iw)}{\Gamma(\frac12 -iw)}.
\label{gamman}
\end{equation}
Expanding $\Upsilon(n, \alpha)$ up to quadratic order in $\alpha$, then Eq.~\eqref{eq:FHalphabeta} 
can be approximated as
\begin{equation}\label{eq:quadratic_non_univ_const}
 c_{n;\alpha,\beta}\approx e^{-\frac{1}{3}\left(n-\frac{1}{n}\right)\ln 2+2\Upsilon(n,0)-\frac{\zeta_n}{2\pi^2 n}(\alpha^2+\beta^2)},
\end{equation}
where $\zeta_n=\ln 2-2\pi^2 n\gamma_2(n)$ and
\begin{equation}
 \gamma_2(n)=\frac{1}{2}\left.\frac{\partial^2\Upsilon(n, \alpha)}{\partial\alpha^2}\right|_{\alpha=0}=
 \frac{n i}{4}\int_0^\infty {\rm d}w [\tanh^3(\pi n w)-\tanh(\pi n w)]\log\frac{\Gamma(\frac{1}{2}+iw)}{\Gamma(\frac{1}{2}-iw)}.
\end{equation}

In the limit of adjacent intervals, given by Eq.~\eqref{eq:Charged_alphabeta_adj}, the 
multiplicative constant $\tilde{c}_{n;\alpha,\beta}$ cannot 
be determined using the known results for Toeplitz determinants. 
However, we can conjecture an analytical approximation for it at quadratic order in $\alpha$ and $\beta$. When $d\to0$, we can associate to each end-point of the intervals $u_1$, $v_1=u_2$ and $v_2$ the fluxes $\alpha$,
$\beta-\alpha$ and $-\beta$ respectively. From the results for one interval of Ref.~\cite{riccarda}, one can conjecture that each 
end-point with flux $\alpha$ contributes with a factor $e^{-\frac{\zeta_n}{4\pi^2n}\alpha^2}$ to the constant $\tilde{c}_{n;\alpha, \beta}$, if we restrict to terms
up to order $\alpha^2$. Therefore, the combination of all 
the fluxes in our case should contribute with a total factor $e^{-\frac{\zeta_n}{4\pi^2n}(\alpha^2+(\beta-\alpha)^2+\beta^2)}$.
We then expect that $\tilde{c}_{n;\alpha,\beta}$ should be well be approximated by
\begin{equation}
\label{eq:FHalphabeta_adj}
\tilde{c}_{n;\alpha,\beta}=e^{-\frac{1}{6}\left(n-\frac{1}{n}\right)\ln 2+\Upsilon(n,0)-\frac{\zeta_n}{2\pi^2 n}(\alpha^2+\beta^2-\alpha\beta)}.
\end{equation}
When $\alpha=\beta=0$, the expression above simplifies to the 
multiplicative constant for the moments of a single interval
obtained in Ref.~\cite{JK04}. In spite of the heuristic reasoning of this result, in Fig.~\ref{fig:Chargedmomentsdnon0}, we check its validity by comparing it against exact numerical data.

In the top and middle panels of Fig.~\ref{fig:Chargedmomentsdnon0}, we study $Z_1^{A_1:A_2}(\alpha,\beta)$ as function of $\alpha$, for various values of $\beta$, $\ell_1$, $\ell_2$ and $d>0$. The points correspond to the exact numerical values obtained as we described in Appendix~\ref{app:numericaltools}, while the solid curves are 
the analytic prediction of Eq.~(\ref{eq:Charged_alphabeta_disj}), 
taking as multiplicative constant $c_{n,\alpha, \beta}$ that in Eq.~(\ref{eq:FHalphabeta}). We find an excellent agreement. In the bottom left panel, we analyse the case of adjacent intervals ($d=0$). The numerical data for $Z_1^{A_1:A_2}(\alpha,\beta)$ are in very good agreement with the analytical expression (solid curves) of Eq.~\eqref{eq:Charged_alphabeta_adj}, using  Eq.~\eqref{eq:FHalphabeta_adj} as multiplicative constant.
Finally, in the bottom right panel, we plot $Z_1^{A_1:A_2}(\alpha,\beta)$ as function of the cross-ratio $x$, for various values of $\alpha$, $\beta$, $\ell_1$ and $\ell_2$. The curves represent the prediction of Eq.~(\ref{eq:Charged_alphabeta_disj}). The continuous ones
correspond to take as non-universal constant $c_{n;\alpha,\beta}$ the full expression of Eq.~\eqref{eq:FHalphabeta}, while for the dashed
ones we have used the quadratic approximation of Eq.~\eqref{eq:quadratic_non_univ_const}.
The agreement between the analytic prediction and the numerical data is extremely good, even considering the quadratic approximation for the 
non-universal constant. As expected, this agreement is better for small values of $\alpha$ and $\beta$, while, around $\pm \pi$, 
we need to take larger subsystem sizes $\ell_1$, $\ell_2$ in order to suppress the finite-size corrections which are well-known and characterised
for the charged moments with a single flux insertion \cite{riccarda}.

\section{Free compact boson}
\label{sec:CompactBoson}

The second theory we focus on in this manuscript is the 
free compact boson, which is the CFT of the Luttinger liquid, and whose action reads
 \begin{equation}
\label{eq:lagrangianboson}
\mathcal{S}_{\rm b}=\frac{1}{8\pi}\int {\rm d}x_0 {\rm d}x_1 \partial_\mu \varphi \partial^{\mu} \varphi.
\end{equation}  
The target space of the real field $\varphi$ is compactified on a circle of radius 
$R$, i.e. $\varphi \sim \varphi +2\pi m R $ with $m\in \mathbb{Z}$. The 
compactification radius $R$ is related to the Luttinger parameter $K$ as $R=\sqrt{2/ K}$.
The action of Eq.~(\ref{eq:lagrangianboson}) is invariant under the transformation 
$\varphi \mapsto \varphi+\alpha $ which, due to the compact nature of $\varphi$, 
realises a $U(1)$ global symmetry. The associated conserved charge is 
$Q_{\rm b}=\frac{1}{2\pi}\int {\rm d}x_1\partial_{x_1} \varphi$. 

The moments of the ground state reduced density matrix of 
this theory are well-known. An exact analytic expression for the 
two-interval case was obtained in Ref.~\cite{twist2}, which was generalised
to an arbitrary number of disjoint intervals in Ref.~\cite{CoserTagliacozzo}.
In particular, for two intervals, it reads~\cite{twist2}
\begin{equation}
\label{eq:compactboson2intCCT}
 Z_n^A(0)=c_{n}\left[\ell_1 \ell_2 (1-x)\right]^{\frac{1-n^2}{6n}}\mathcal{F}_n(x),
\end{equation} 
where $c_{n}$ is a non-universal constant and
\begin{equation}\label{eq:mathcal_F}
 \mathcal{F}_n(x)=\frac{\Theta(\boldsymbol{0}|\Gamma(x)/K)\Theta (\boldsymbol{0}|\Gamma(x)K)}{\left[\Theta (\boldsymbol{0}|\Gamma(x))\right]^2}.
\end{equation}
We denote by $\Theta$ the Riemann-Siegel Theta function 
\begin{equation}
\label{eq:Siegel_Theta}
\chartheta{\boldsymbol{\varepsilon}}{\boldsymbol{\delta}} (\boldsymbol{u}|\Omega)\equiv
\sum_{\boldsymbol{m}\in\mathbb{Z}^{n-1}}
e^{i\pi(\boldsymbol{m}+\boldsymbol{\varepsilon})^t
\cdot \Omega(\boldsymbol{m}+\boldsymbol{\varepsilon})+2\pi i 
(\boldsymbol{m}+\boldsymbol{\varepsilon})^t\cdot(\boldsymbol{u}+\boldsymbol{\delta})},
\end{equation}
with characteristics $\boldsymbol{\varepsilon},\boldsymbol{\delta}\in
(\mathbb{Z}/2)^{n-1}$, $\boldsymbol{u}\in\mathbb{C}^{n-1}$ 
and $\Omega$ a complex $(n-1)\times(n-1)$ matrix. In Eq.~\eqref{eq:mathcal_F},
the characteristics are zero $\boldsymbol{0}=(0,\dots,0)$ and, therefore, we
have used the standard shorthand notation $\Theta(\boldsymbol{u}|\Omega)\equiv 
\chartheta{\boldsymbol{0}}{\boldsymbol{0}} (\boldsymbol{u}|\Omega)$.
The matrix $\Gamma(x)$ in Eq.~(\ref{eq:mathcal_F}) 
has entries given by
\begin{equation}\label{eq:matrix_periods}
\Gamma_{rs}(x)=\frac{2i}{n}
\sum_{l=1}^{n-1}\cos\left[\frac{2 \pi l (r-s)}{n}\right]\sin\left(\frac{\pi l}{n}\right)
\beta_{l/n}(x),\quad r, s=1, \dots,n-1
,
\end{equation}
and
\begin{equation}
\label{eq:beta_def}
\beta_p(x)=\frac{I_p(1-x)}{I_p(x)},
\end{equation}
with $I_p(x)\equiv \,_2F_1(p,1-p,1,1-x)$. The function
$\mathcal{F}_n(x)$ is invariant under $x\mapsto 1-x$ and it is
normalised such that $\mathcal{F}_n(0)=\mathcal{F}_n(1)=1$.
Although the moments of $\rho_A$ are known for all the integer $n$, 
its analytic continuation to complex $n$ and, consequently, the 
von Neumann entropy of Eq.~\eqref{eq:defEE} is still not available 
for all the values of the Luttinger parameter.

A remarkable contact point between the theories described by the 
actions of Eqs.~\eqref{eq:lagrangianDirac} and \eqref{eq:lagrangianboson} is the case $K=1$. Notice that, when the
Luttinger parameter takes this value, the function $\mathcal{F}_n(x)$
in Eq.~\eqref{eq:mathcal_F} simplifies to $\mathcal{F}_n(x)$=1 and the
moments of Eq.~\eqref{eq:compactboson2intCCT} for the massless compact
boson present the same universal dependence on $\ell_1$, $\ell_2$ and
$x$ as the ones in Eq.~\eqref{eq:Dirac2intCFH} for the massless Dirac
fermion. A detailed discussion on this identity can be found in 
Ref.~\cite{HLM13}, where it is explained the reason why, although
these two theories are not related by a duality, their partition 
functions on the Riemann surfaces $\Sigma_n$ arising in the 
two-interval replica method are actually equal. 
Here we find that this identity extends to the partition
functions $Z_n^{A_1:A_2}(\{\alpha_j\})$ on the surface $\Sigma_n$
with different twisted boundary conditions at each branch point.
In general,
when $A$ is made up of more than two intervals and the R\'enyi
index $n$ is larger than two, the moments of the reduced density
matrix in these CFTs (and the corresponding R\'enyi entropies) 
are different~\cite{HLM13}.

\subsection{Charged moments}
We now generalise the result~\eqref{eq:compactboson2intCCT} to the multi-charged moments in Eq.~\eqref{eq:Charged_alphabeta}. 
Starting from  Eq.~\eqref{eq:gen_charged_mom_corr_riemann}, we will compute them as the four-point function of the field 
$\mathcal{V}_\alpha$ on the Riemann surface $\Sigma_n$. 
Since the $U(1)$ conserved current is proportional to $\partial_{x_1} \varphi$, $\mathcal{V}_\alpha$ 
is identified in this case with the vertex operator \cite{goldstein}
\begin{equation}\label{eq:vertex_op}
 \mathcal{V}_\alpha(z)=e^{i\frac{\alpha}{2\pi}\varphi(z)},
\end{equation}
which has conformal dimensions
\begin{equation}
 h_\alpha^{\mathcal{V}}=\bar{h}_\alpha^{\mathcal{V}}=\left(\frac{\alpha}{2\pi}\right)^2\frac{K}{2}.
\end{equation}
In the following, it will be useful to introduce the \textit{rescaled} Luttinger paratemeter $\eta=K/(2\pi^2)$ in order to lighten the expressions.

Without loss of generality, let us consider that the end-points of subsystem $A$ are $u_1=0$, $v_1=x$, 
$u_2=1$ and $v_2=\infty$. Using Eq.~\eqref{eq:twist_gen_partition_func}, 
and given that the composite twist fields are primaries, 
we can eventually obtain the expression for an arbitrary set of end-points 
through a global conformal transformation. Therefore, according to Eq.~\eqref{eq:gen_charged_mom_corr_riemann}, 
the multi-charged  moments can be derived from the four-point correlation 
function of the vertex operators of Eq~\eqref{eq:vertex_op}
\begin{equation}\label{eq:gen_bos_charged_mom_riemann}
 Z_n^{A_1:A_2}(\{\alpha_j\})=Z_n^{A}(0)\langle \mathcal{V}_{\alpha_1}(0)
 \mathcal{V}_{\alpha_2}(x)\mathcal{V}_{\alpha_3}(1)\mathcal{V}_{\alpha_4}(\infty)\rangle_{\Sigma_{n}(x)}
\end{equation}
on the $n$-sheeted Riemann surface $\Sigma_n(x)$ with branch points 
at $0$, $x$, $1$ and $\infty$. This surface of genus $n-1$ can be described by the complex curve
\begin{equation}\label{eq:algebraic_curve}
 y^n=\frac{z(z-1)}{z-x}.
\end{equation}
The correlator of vertex operators on a general Riemann surface 
of arbitrary genus  was obtained in Ref.~\cite{Verlinde}. In order 
to give the explicit expression in our case, we need to introduce 
some notions about Riemann surfaces~\cite{Fay}.

There are different parameterisations of the moduli space of genus $n-1$ Riemann surfaces. 
One possibility is through the matrix of periods, which we denote by $\Gamma$. This 
is a $(n-1)\times (n-1)$ symmetric matrix with positive definite imaginary part. Notice that, according to Eq.~\eqref{eq:algebraic_curve},
the Riemann surface $\Sigma_n(x)$ is parametrised by the cross-ratio $x$. 
Therefore, the corresponding matrix of periods only depends on $x$, i.e. $\Gamma=\Gamma(x)$. 
In order to define it, we need first to specify a particular homology 
basis for $\Sigma_n(x)$, i.e. a basis of $2(n-1)$ oriented non-contractible curves on the surface, which we denote 
by $a_r$ and $b_r$, with $r=1, \dots, n-1$. The detailed description of the specific basis that we consider is given in Appendix~\ref{app:riemann}. We also have to choose a basis of holomorphic 
differentials $\nu_r$, $r=1, \dots, n-1$, normalised with respect to the $a_r$ cycles. 
That is,
\begin{equation}\label{eq:norm_cond_nu}
 \oint_{a_r}{\rm d}z \nu_s(z)=\delta_{r,s},\quad r,s=1, \dots, n-1.
\end{equation}
Then the matrix of periods is defined as
\begin{equation}\label{eq:def_matrix_periods}
 \Gamma_{rs}=\oint_{b_r}{\rm d}z \nu_s(z).
\end{equation}
For the surface $\Sigma_n(x)$, the normalised holomorphic differentials
read
\begin{equation}\label{eq:norm_diff}
 \nu_r(z)=\frac{1}{\pi n}\sum_{l=1}^{n-1}
 \frac{e^{-\frac{i 2\pi (r-1)l}{n}}\sin(\pi l/n)}{I_{l/n}(x)}
 (z(z-1))^{-l/n}(z-x)^{-1+l/n}.
\end{equation}
In Appendix~\ref{app:riemann}, we thoroughly explain the derivation 
of 
the expression for $\nu_r$. Inserting it in Eq.~\eqref{eq:def_matrix_periods}, it is then easy to show that the 
entries of the matrix of periods $\Gamma(x)$ are precisely those of Eq.~\eqref{eq:matrix_periods}.

If we now consider four vertex 
operators inserted at generic points in the surface $\Sigma_n(x)$ and with arbitrary 
dimensions satisfying the neutrality condition $\alpha_1+\alpha_2+\alpha_3+\alpha_4=0$,
then its correlation function is of the form~\cite{Verlinde}
\begin{multline}\label{eq:general_vertex_corr_genus_g}
 \langle \mathcal{V}_{\alpha_1}(z_1) \mathcal{V}_{\alpha_2}(z_2) 
 \mathcal{V}_{\alpha_3}(z_3) \mathcal{V}_{\alpha_4}(z_4)\rangle_{\Sigma_n(x)}=\\
 \prod_{1\leq j < j'\leq 4} \left|E(z_j, z_{j'})e^{-\pi {\rm Im}[\boldsymbol{w}(z_j)-\boldsymbol{w}(z_{j'})]^{t}\cdot{\rm Im}[\Gamma(x)^{-1}]
 {\rm Im}[\boldsymbol{w}(z_j)-\boldsymbol{w}(z_{j'})]}\right|^{\alpha_j\alpha_{j'}\eta}.
\end{multline}
In this expression, we denote by $E(z, z')$ the prime form of the surface $\Sigma_n(x)$,
 which we will define precisely later,  
and $\boldsymbol{w}(z)=(w_1(z), \dots, w_{n-1}(z))$ 
is the Abel-Jacobi map, which relates a point $z$ in the surface $\Sigma_n(x)$ to a point $\boldsymbol{w}(z)$ 
in the genus $n-1$ complex torus $\mathbb{C}^{n-1}/\Lambda$, where $\Lambda=\mathbb{Z}^{n-1}+\Gamma\mathbb{Z}^{n-1}$.
This map can be written in terms of the normalised holomorphic differentials of
Eq.~\eqref{eq:norm_diff} as
\begin{equation}
 w_r(z)=\int_0^z {\rm d} z' \nu_r(z')\quad (\rm{mod}\,\Lambda),
\end{equation}
where we have taken as origin the branch point $z=0$.
The images under the Abel-Jacobi map of the points $z=0$, $x$, $1$, and $\infty$, 
where the vertex operators in Eq.~\eqref{eq:gen_bos_charged_mom_riemann} are 
inserted, can be easily computed using Eq.~\eqref{eq:norm_diff} and 
applying the identities of Eq.~\eqref{eq:int_identities}. Then we find
\begin{eqnarray}
 \boldsymbol{w}(0)&=&\boldsymbol{0},\label{eq:abel_map_1}\\
 \boldsymbol{w}(x)&=&\boldsymbol{q},\label{eq:abel_map_2}\\
 \boldsymbol{w}(1)&=&\boldsymbol{q}+i\boldsymbol{p}(x),\label{eq:abel_map_3}\\
 \boldsymbol{w}(\infty)&=&i\boldsymbol{p}(x),\label{eq:abel_map_4}
\end{eqnarray}
where $\boldsymbol{q}=(1/n, \dots,1/n)$ and $\boldsymbol{p}(x)=(p_1(x), \dots, p_{n-1}(x))$
with
\begin{equation}
 p_r(x)=-\frac{1}{n}\sum_{l=1}^{n-1}
 \left[\cos\left[\frac{2\pi l(r-1)}{n}\right]\sin\left(\frac{\pi l}{n}\right)
 +\sin\left[\frac{2\pi l(r-1)}{n}\right]\cos\left(\frac{\pi l}{n}\right)\right]\beta_{l/n}(x).
\end{equation}
Therefore, for the case $z_1=0$, $z_2=x$, $z_3=1$, $z_4=\infty$, 
Eq.~\eqref{eq:general_vertex_corr_genus_g} simplifies to
\begin{equation}\label{eq:simp_v_corr_funct}
 \langle \mathcal{V}_{\alpha_1}(0) \mathcal{V}_{\alpha_2}(x) 
 \mathcal{V}_{\alpha_3}(1) \mathcal{V}_{\alpha_4}(\infty)\rangle_{\Sigma_n(x)}=
M_n(x)^{(\alpha_1+\alpha_2)(\alpha_3+\alpha_4)\eta}
\prod_{1\leq j < j'\leq 4} \left|E(z_j, z_{j'})\right|^{\alpha_{j}\alpha_{j'}\eta},
\end{equation}
where
\begin{equation}
 M_n(x)=e^{-\pi \boldsymbol{p}(x)^t
\cdot [{\rm Im}\Gamma(x)]^{-1}\boldsymbol{p}(x)}. 
\end{equation}

Let us now focus on the prime form $E(z, z')$. It 
can be defined as~\cite{Fay}
\begin{equation}
 E(z, z')=\frac{\Theta_{\boldsymbol{\frac{1}{2}}}
 (\boldsymbol{w}(z)-\boldsymbol{w}(z')|\Gamma(x))}
 {\sqrt{g(z)}\sqrt{g(z')}},
\end{equation}
where $\Theta_{\boldsymbol{\frac{1}{2}}}$ is a shorthand notation for 
the Theta function of Eq.~\eqref{eq:Siegel_Theta} with both characteristics equal to 
$(1/2, 0,\dots, 0)\in(\mathbb{Z}/2)^{n-1}$ and $g(z)$ is 
\begin{equation}
 g(z)=\sum_{r=1}^{n-1} \nu_r(z)\partial_{u_r}
 \left.\Theta_{\boldsymbol{\frac{1}{2}}}(\boldsymbol{u}|\Gamma(x))\right|_{\boldsymbol{u}=\boldsymbol{0}}.
\end{equation}

Notice  that the holomorphic differentials $\nu_r(z)$ in Eq.~\eqref{eq:norm_diff} and, therefore, $g(z)$ are singular
at the branch points of the curve that defines the surface $\Sigma_n(x)$. 
This means that the correlation function~\eqref{eq:general_vertex_corr_genus_g} is in principle not well-defined. 
In order to solve this issue, the vertex operators inserted at the branch points
have to be regularised by redefining them as a proper limit from a non-singular point. 
We will first extract and remove from $g(z)$ the divergent terms at the branch points. 
Then, by considering the limit in which the distance between $A_1$ and $A_2$ tends to
infinity, we will fix the correct definition of
the regularised vertex operators at the branch points.

Close to the branch points $z=0$, $x$, $1$, and $\infty$, the holomorphic 
normalised differentials $\nu_r(z)$ of Eq.~\eqref{eq:norm_diff} behave as
\begin{equation}
 \nu_r(z+\epsilon)=\epsilon^{\frac{1-n}{n}}\left[\nu_r^{(*)}(z)+O(\epsilon^{1/n})\right],
\end{equation}
with $|\epsilon|\ll 1$ , 
\begin{equation}\label{eq:reg_norm_hol_diff}
 \nu_r^{(*)}(z)=\left\{\begin{array}{ll}
 x^{-1/n}Q_{r,n}(x),& z=0,\\
 e^{-\frac{i\pi(4r-3)}{n}}(x(1-x))^{-1/n}Q_{r, n}(x),& z=x,\\
 (1-x)^{-1/n}Q_{r, n}(x),& z=1,\\
 e^{-\frac{4\pi i(r-1)}{n}}Q_{r, n}(x),& z=\infty,
 \end{array}\right.
\end{equation}
and 
\begin{equation}
 Q_{r,n}(x)=e^{\frac{2\pi i(r-1)}{n}}\frac{\sin(\pi/n)}{\pi n I_{1/n}(x)}.
\end{equation}
Observe that, in the four singularities, the divergent term when 
$\epsilon\to 0$  is a global factor $\epsilon^{\frac{1-n}{n}}$ and, once we take it out, the 
subleading corrections in $\epsilon$ vanish. Therefore, these singularities can be 
removed in the correlation function of 
Eq~\eqref{eq:general_vertex_corr_genus_g} 
by defining the vertex operators at the branch points as the limit 
\begin{equation}\label{eq:reg_vertex}
 \mathcal{V}_\alpha^{(*)}(z)=
 \lim_{\epsilon\to 0}\left(\kappa_n \epsilon^{\frac{n-1}{n}}\right)^{2h_\alpha^{\mathcal{V}}} 
 \mathcal{V}_\alpha(z+\epsilon),
 \quad z=0, x, 1, \infty.
\end{equation}
In this definition, we have included a possible global rescaling 
factor $\kappa_n$, which may depend on the genus of the surface, and 
we will adjust by studying the limit of large separation between the 
two intervals. If we replace in Eq.~\eqref{eq:simp_v_corr_funct} the 
vertex operators by the regularised ones introduced in 
Eq.~\eqref{eq:reg_vertex}, then the resulting correlation function 
can be written in the form
\begin{multline}\label{eq:reg_v_corr_funct}
 \langle \mathcal{V}_{\alpha_1}^{(*)}(0)\mathcal{V}_{\alpha_2}^{(*)}(x)
 \mathcal{V}_{\alpha_3}^{(*)}(1)\mathcal{V}_{\alpha_4}^{(*)}(\infty)\rangle_{\Sigma_n(x)}=\\=
 \kappa_n^{2h_T} M_n(x)^{(\alpha_1+\alpha_2)(\alpha_3+\alpha_4)\eta}
\prod_{1\leq j<j'\leq 4}|E^{(*)}(z_j, z_{j'})|^{\alpha_j\alpha_{j'}\eta},
\end{multline}
where $h_T=h_{\alpha_1}+h_{\alpha_2}+h_{\alpha_3}+h_{\alpha_4}$ 
and $E^{(*)}(z_j, z_j')$ stands for the regularised prime form 
\begin{equation}\label{eq:reg_prime_form}
 E^{(*)}(z_j, z_{j'})=\frac{\Theta_{\boldsymbol{\frac{1}{2}}}(\boldsymbol{w}(z_j)-\boldsymbol{w}(z_{j'})|\Gamma(x))}
 {\sqrt{g^{(*)}(z_j)}\sqrt{g^{(*)}(z_{j'})}},
\end{equation}
with
\begin{equation}
 g^{(*)}(z_j)=\sum_{r=1}^{n-1}\nu_r^{(*)}(z_j)\partial_{u_r}\left.\Theta_{\boldsymbol{\frac{1}{2}}}(\boldsymbol{u}|\Gamma(x))\right|_{\boldsymbol{u}=\boldsymbol{0}},
\end{equation}
and the expressions of $\nu^{(*)}(z)$ at the branch points are those 
given in Eq.~\eqref{eq:reg_norm_hol_diff}.

In Appendix~\ref{app:prime_forms}, we conjecture and numerically check the following identities
for the regularised prime forms that appear in Eq.~\eqref{eq:reg_v_corr_funct},
\begin{eqnarray}
 |E^{(*)}(0, x)|&=& n x^{1/n},\label{eq:prime_form_id_1}\\
 |E^{(*)}(x, 1)|&=& \frac{n (1-x)^{1/n}}{M_n(x)},\label{eq:prime_form_id_2}\\
 |E^{(*)}(1, \infty)|&=& n,\label{eq:prime_form_id_3}
\end{eqnarray}
and
\begin{equation}\label{eq:prime_form_id_4}
 |E^{(*)}(0, 1)|=|E^{(*)}(0,\infty)|=|E^{(*)}(x,\infty)|=\frac{n}{M_n(x)}.
\end{equation}
Plugging them into Eq.~\eqref{eq:reg_v_corr_funct}, we finally find 
\begin{equation}\label{eq:final_reg_v_corr_funct}
 \langle \mathcal{V}_{\alpha_1}^{(*)}(0)\mathcal{V}_{\alpha_2}^{(*)}(x) 
 \mathcal{V}_{\alpha_3}^{(*)}(1) \mathcal{V}_{\alpha_4}^{(*)}(\infty)\rangle_{\Sigma_{n}(x)}
 = \left(\frac{\kappa_n}{n}\right)^{2h_T} x^{\frac{\alpha_1\alpha_2 \eta}{n}}(1-x)^{\frac{\alpha_2\alpha_3 \eta}{n}}.
\end{equation}
In Eq.~\eqref{eq:gen_bos_charged_mom_riemann}, once the vertex operators $\mathcal{V}_\alpha$ are replaced by the regularised ones $\mathcal{V}_\alpha^{(*)}$, we 
can exploit Eq.~\eqref{eq:final_reg_v_corr_funct} to get
\begin{equation}
 Z_n^{A_1:A_2}(\{\alpha_j\})\propto
 \left(\frac{\kappa_n}{n}\right)^{2h_T}
 \left(x(1-x)\right)^{\frac{1-n^2}{6n}}\left(x^{\alpha_1\alpha_2}(1-x)^{\alpha_2\alpha_3}\right)^{\frac{\eta}{n}}
 \mathcal{F}_n(x).
\end{equation}
In particular, when $\alpha_1=-\alpha_2=\alpha$ and
$\alpha_3=-\alpha_4=\beta$, we get the multi-charged moments $Z_n^{A_1:A_2}(\alpha,\beta)$. After a global conformal transformation
to a subsystem $A$ with arbitrary end-points $(u_1, v_1, u_2, v_2)$, we obtain the following result
\begin{equation}\label{eq:bos_charged_mom_kappa}
 Z_n^{A_1:A_2}(\alpha,\beta)=c_{n;\alpha,\beta}\left(\frac{\kappa_n}{n}\right)^{2h_T}
 \left(\ell_1\ell_2(1-x)\right)^{\frac{1-n^2}{6n}}
 \left(\ell_1^{\alpha^2}\ell_2^{\beta^2}(1-x)^{\alpha\beta}\right)^{-\frac{\eta}{n}}
 \mathcal{F}_n(x).
\end{equation}

Note that the rescaling factor $\kappa_n$, which was introduced in 
the definition of the regularised vertex operators at the branch 
points, is still undetermined. We can fix it by analysing the limit 
in which the two intervals $A_1$ and $A_2$ are far, i.e. $d\to \infty$, as done for the Dirac theory. 
In that case, the charged moments $Z_n^{A_1:A_2}(\alpha,\beta)$ must verify Eq.~\eqref{eq:far_intervals}.
Since $\mathcal{F}_n(0)=1$ and the constant $c_{n;\alpha,\beta}$ 
factorises into those for the intervals $A_1$ and $A_2$, then Eq.~\eqref{eq:bos_charged_mom_kappa} satisfies the limit $d\to\infty$
of Eq.~\eqref{eq:far_intervals} if $\kappa_n=n$. 

In conclusion, for the massless compact boson, the partition function on the surface $\Sigma_n$ with general twisted boundary 
conditions is of the form
\begin{equation}
 \frac{Z_n^{A_1:A_2}(\{\alpha_j\})}{Z_n^{A}(0)}\propto
\left[d^{\alpha_2 \alpha_3}
\ell_1^{\alpha_1 \alpha_2}
\ell_2^{\alpha_3 \alpha_4}
(d+\ell_1)^{\alpha_1 \alpha_3}
(d+\ell_2)^{\alpha_2 \alpha_4}
(d+\ell_1+\ell_2)^{\alpha_1 \alpha_4}\right]^{\frac{K}{2\pi^2 n}},
\end{equation}
and the multi-charged moments are
\begin{equation}\label{eq:bos_charged_mom_final}
 Z_n^{A_1:A_2}(\alpha,\beta)=c_{n;\alpha,\beta}
 \left(\ell_1\ell_2(1-x)\right)^{\frac{1-n^2}{6n}}
 \left(\ell_1^{\alpha^2}\ell_2^{\beta^2}(1-x)^{\alpha\beta}\right)^{-\frac{K}{2\pi^2n}}
 \mathcal{F}_n(x).
\end{equation}
When the Luttinger parameter is $K=1$, then $\mathcal{F}_n(x)=1$, and the partition function of Eq.~\eqref{eq:bos_charged_mom_final} is equal to the one obtained in Eq.~\eqref{eq:ChargedMom4fluxestotal_v2} for the massless Dirac field, as we anticipated at the beginning of this section.
Interestingly, the factor in Eq.~\eqref{eq:bos_charged_mom_final} due to the magnetic
fluxes is the same, when $\alpha=\beta$, as the one derived in Ref.~\cite{wznm-21}
for a large central charge CFT with the Luttinger parameter $K$ replaced by the level of the Kac-Moody algebra of that theory.

\section{Symmetry resolution}
\label{sec:SymmetryResolution}

In this section, we apply the approach described in Sec.~\ref{sub:symi} in order to evaluate the symmetry resolution of the mutual information in the two CFTs analysed in Secs.~\ref{sec:FreeDirac}
and \ref{sec:CompactBoson} from the expressions obtained there for their multi-charged
moments $Z_n^{A_1:A_2}(\alpha,\beta)$.

\subsection{Fourier transforms}\label{subsec:Dirac_qmom_adjacent}
The first step is to determine the Fourier transform \eqref{eq:Znq_alphabeta} of the multi-charged moments. 
We need to know how the non-universal constant $c_{n;\alpha,\beta}$ does depend on $\alpha$ and $\beta$. 
In Sec.~\ref{sec:FreeDirac}, we have concluded that, for
the tight-binding model, it can be well approximated if we only take into account the quadratic terms in $\alpha$ and $\beta$. In the following, we will assume that this is in general a good approximation ~\cite{xavier}. Therefore, we will take 
\begin{equation}\label{eq:quadratic_non_univ_const_2}
c_{n;\alpha,\beta}=c_{n;0,0}\lambda_{n}^{-\frac{(\alpha^2+\beta^2)K}{2\pi^2 n}}.
\end{equation}
In the case of the tight-binding model ($K=1$), we obtained in
Eq.~\eqref{eq:quadratic_non_univ_const} that $\lambda_{n}=e^{\zeta_n}$.

Therefore, applying Eq.~\eqref{eq:quadratic_non_univ_const_2} in the result of Eq.~\eqref{eq:bos_charged_mom_final}, 
the multi-charged moments can be rewritten as
\begin{equation}\label{eq:gen_charged_mom_quadratic}
Z_n^{A_1:A_2}(\alpha,\beta)=Z_n^{A}(0)\left[\left(1-x\right)^{-\alpha \beta}
\tilde{\ell}_1^{-\alpha^2 }
\tilde{\ell}_2^{-\beta^2}
\right]^{\frac{K}{2\pi^2 n}},
\end{equation}
where $\tilde{\ell}_p=\lambda_{n}\ell_p$.
The evaluation of Eq. \eqref{eq:Znq_alphabeta} using the expression above yields the following multivariate Gaussian function for the Fourier modes of the multi-charged moments
\begin{equation}
\label{eq:Znq_alphabeta_disj}
\mathcal{Z}_n^{A_1:A_2}(q_1,q_2)= \frac {Z_n^{A}(0)n\pi e^{- 2 \pi^2 n\frac{q_1^2\ln\tilde{\ell}_2+q_2^2\ln \tilde{\ell}_1+
q_1q_2 \ln(1-x)}{K[4\ln(\tilde{\ell}_1)\ln(\tilde{\ell}_2)-\ln^2(1-x)]}} }{K\sqrt{4\ln(\tilde{\ell}_1)\ln(\tilde{\ell}_2)-\ln^2(1-x)}}.
\end{equation}
Notice  that the Luttinger parameter $K$ enters in the Gaussian factor as an overall rescaling of its variance.

In the limit of large separation between the intervals, i.e. $d\to \infty$ ($x\to0$), Eq.~\eqref{eq:Znq_alphabeta_disj} tends to
\begin{equation}
\lim_{d\to\infty}\frac{\mathcal{Z}_n^{A_1:A_2}(q_1,q_2)}{Z_n^A(0)}=\frac{n\pi}{2K}
\frac{e^{-\frac{ n\pi^2q_1^2}{2K\ln\tilde{\ell}_1}}}{\sqrt{\ln\tilde{\ell}_1}}
\frac{e^{-\frac{ n\pi^2q_2^2}{2K\ln\tilde{\ell}_2}}}{\sqrt{\ln\tilde{\ell}_2}},
\end{equation}
namely $\mathcal{Z}_n^{A_1:A_2}(q_1,q_2)$ factorises into the contributions of $A_1$ and $A_2$. 
This is consistent with the probabilistic interpretation for the case $n=1$: the outcomes of 
the charge measurements in the two intervals are independently distributed when the 
separation between $A_1$ and $A_2$ is large enough.
On the other hand, in the limit of two adjacent intervals, i.e. $d\to0$ ($x\to 1$), the multi-charged moments have the form (see also
Eq.~\eqref{eq:Charged_alphabeta_adj})
\begin{equation}
\lim_{d\to 0} \frac{Z_n^{A_1:A_2}(\alpha,\beta)}{Z_n^A(0)}=
\left[\frac{\tilde{\ell}_1^{\,\alpha \beta-\alpha^2 }
\tilde{\ell}_2^{\,\alpha \beta-\beta^2}}{(\tilde{\ell}_1+\tilde{\ell}_2)^{\alpha \beta}}
\right]^{\frac{K}{2\pi^2 n}},
\end{equation}
whose Fourier transform is
\begin{equation}
\label{eq:Znq_alphabeta_adj}
\lim_{d\to 0}\frac{\mathcal{Z}_n^{A_1:A_2}(q_1,q_2)}{Z_n^{A_1\cup A_2}(0)}=
\frac{n\pi e^{-2\pi^2 n\frac{q_1^2\ln\tilde{\ell}_2+q_2^2\ln\tilde{\ell}_1+
q_1q_2 \left[\ln(\tilde{\ell}_1\tilde{\ell}_2)-\ln\left(\tilde{\ell}_1+\tilde{\ell}_2)\right)\right]}
{4K\ln(\tilde{\ell}_1)\ln(\tilde{\ell}_2)-K\left[\ln(\tilde{\ell}_1\tilde{\ell}_2)-\ln\right(\tilde{\ell}_1
+\tilde{\ell}_2)\left)\right]^2}}}{K\sqrt{4\ln(\tilde{\ell}_1)\ln(\tilde{\ell}_2)
-\left[\ln(\tilde{\ell}_1\tilde{\ell}_2)-\ln(\tilde{\ell}_1+\tilde{\ell}_2)\right]^2}}.
\end{equation}

Setting $\alpha=\beta$ in Eq.~\eqref{eq:gen_charged_mom_quadratic}, 
we obtain the charged moments~(\ref{eq:Charged_alpha}) with a single flux
\begin{equation}
Z_n^{A}(\alpha)
=Z_n^{A}(0)\left[(1-x)\tilde{\ell}_1\tilde{\ell}_2\right]^{-\frac{\alpha^2K}{2\pi^2 n}}.
\end{equation}
In this case, performing the Fourier transform of Eq.~(\ref{eq:Znq_alpha}), we end up with
\begin{equation}\label{eq:fourier_tranf_non_charg_mom}
\mathcal{Z}_n^{A}(q)=
\frac{Z_n^{A}(0)\sqrt{\pi n}}{\sqrt{2 K\ln\left[(1-x)\tilde{\ell}_1\tilde{\ell}_2\right]}}
e^{-\frac{\pi^2 n q^2}{2K \ln\left[(1-x)\tilde{\ell}_1\tilde{\ell}_2\right]}}.
\end{equation}
Taking $n=1$ in this expression, we obtain the probability $p(q)$
of having charge $q$ in the subsystem $A$, namely $p(q)=\mathcal{Z}_1(q)$. 
Now we can plug it together with the result for $\mathcal{Z}^{A_1:A_2}_n(q_1, q_2)$ found in Eq.~(\ref{eq:Znq_alphabeta_disj}) into Eq. (\ref{eq:CondProbability}) to obtain the conditional probability $p(q_1, q_2)$ of having charge $q_1$ and $q_2=q-q_1$ in the intervals $A_1$ and $A_2$ if the total charge in $A$ is $q$,
\begin{equation}\label{eq:cond_prob_final}
p(q_1, q_2)= \sqrt{\frac{2\pi\ln\left[(1-x)\tilde{\ell}_1\tilde{\ell}_2\right]}{K[4\ln(\tilde{\ell}_1)\ln(\tilde{\ell}_2)-\ln^2(1-x)]}}
e^{-\frac{2\pi^2}{K}\left[\frac{q_1^2\ln\tilde{\ell}_2+q_2^2\ln \tilde{\ell}_1+
q_1q_2 \ln(1-x)}{4\ln(\tilde{\ell}_1)\ln(\tilde{\ell}_2)-\ln^2(1-x)}+\frac{(q_1+q_2)^2}{4 \ln\left[(1-x)\tilde{\ell}_1\tilde{\ell}_2\right]}\right]}.\end{equation}
In this expression, $\tilde{\ell}_p=\lambda_1\ell_p$; in particular, for the tight-binding model $\lambda_{1}=e^{\ln 2 + 1 + \gamma_{\rm E}}$,
with $\gamma_{\rm E}$ the Euler-Mascheroni constant.
As a non-trivial consistency check, we have verified that the probability functions we have obtained satisfy 
the normalisation conditions 
\begin{equation}\label{eq:normalisation}
\int_{-\infty}^{\infty}\mathcal{Z}_1^{A_1:A_2}(q_1,q_2)\mathrm{d}q_1\mathrm{d}q_2=1,\quad \int_{-\infty}^{q}p(q_1,q-q_1)\mathrm{d}q_1=1,
\end{equation}
in agreement with Eqs. \eqref{eq:CondProbability} and \eqref{eq:normalisationq1q2}.

\begin{figure}[t]
\centering
\subfigure
{\includegraphics[width=0.32\textwidth]{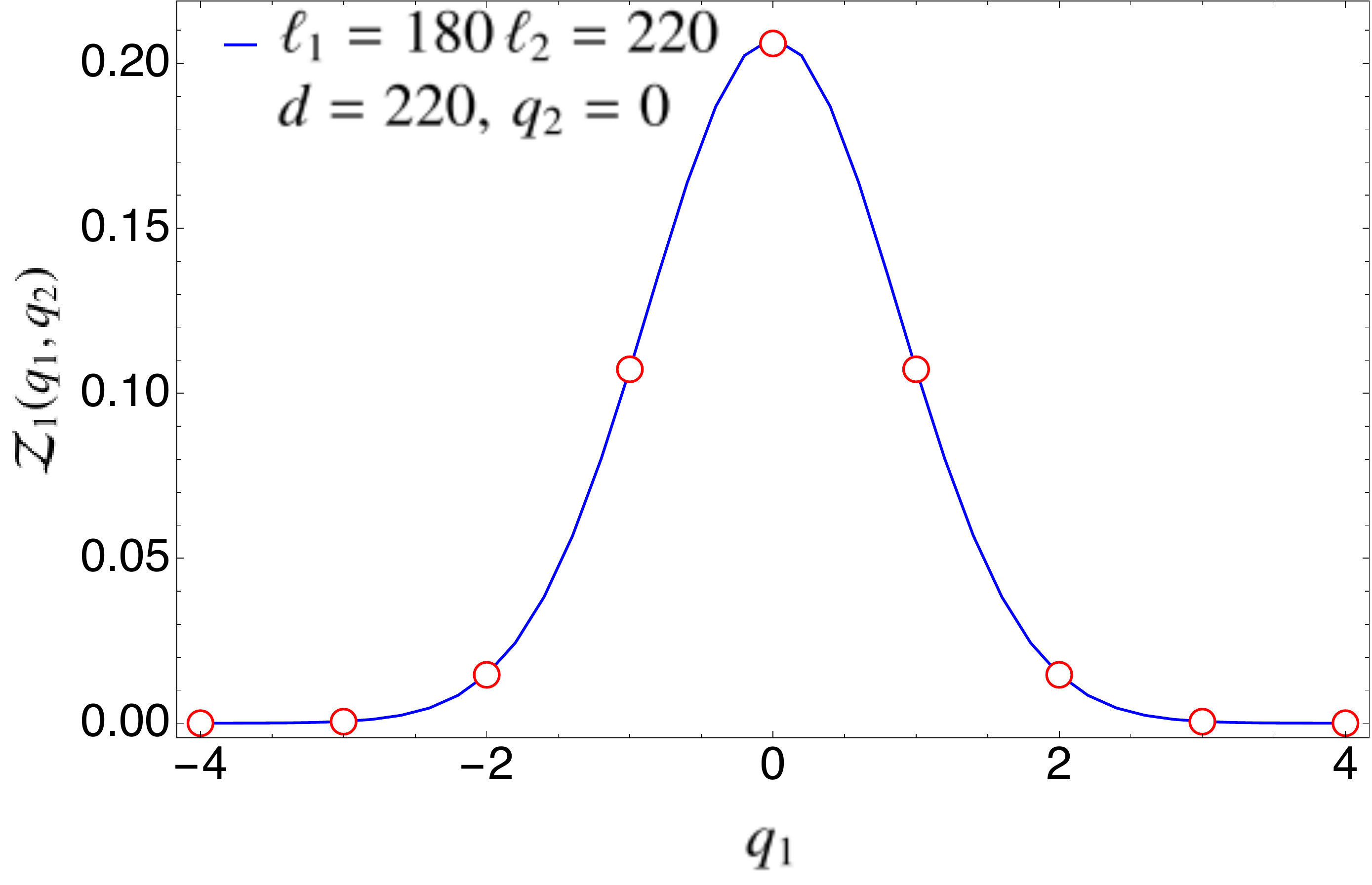}}
\subfigure
{\includegraphics[width=0.32\textwidth]{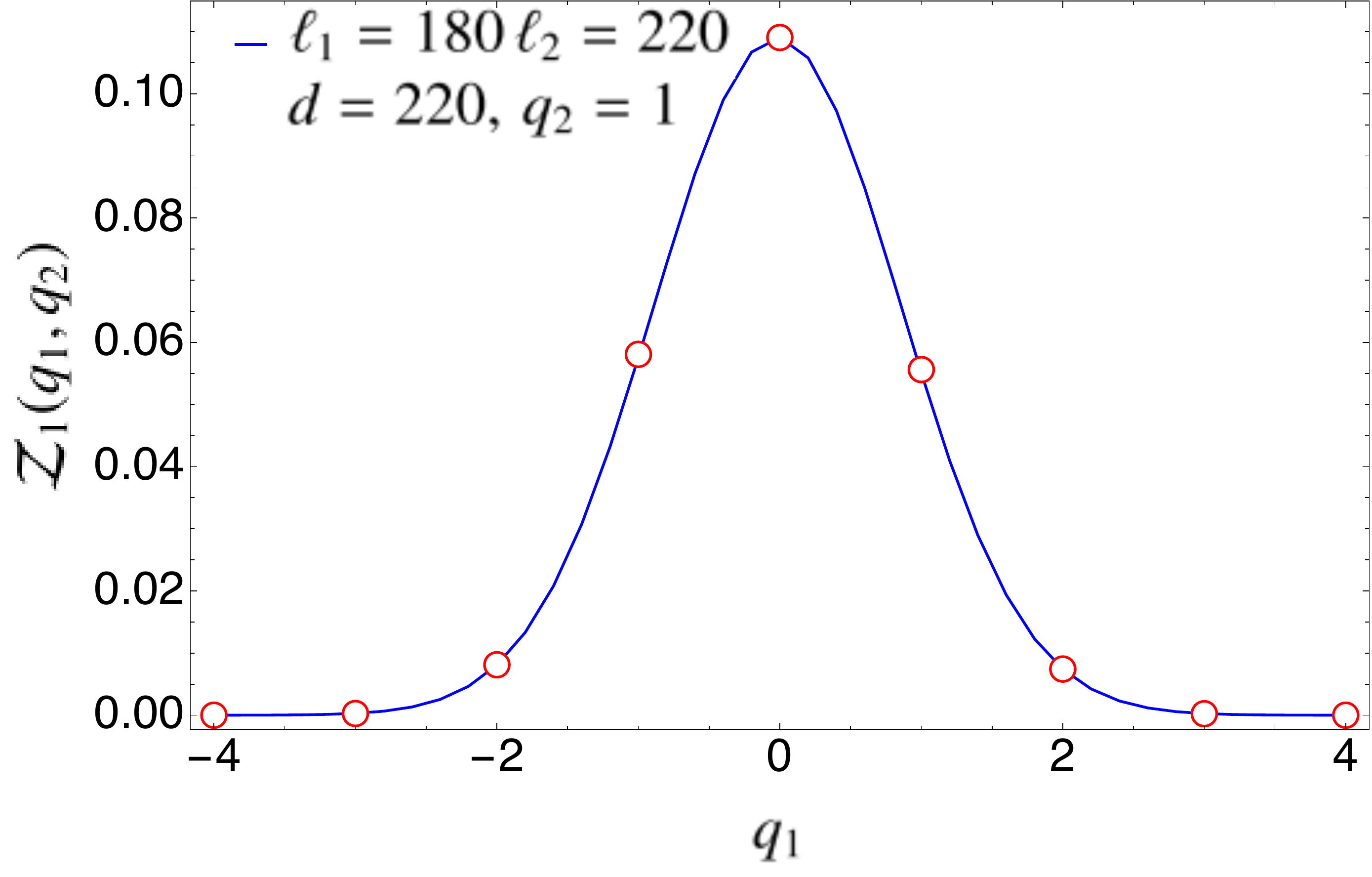}}
\subfigure
{\includegraphics[width=0.32\textwidth]{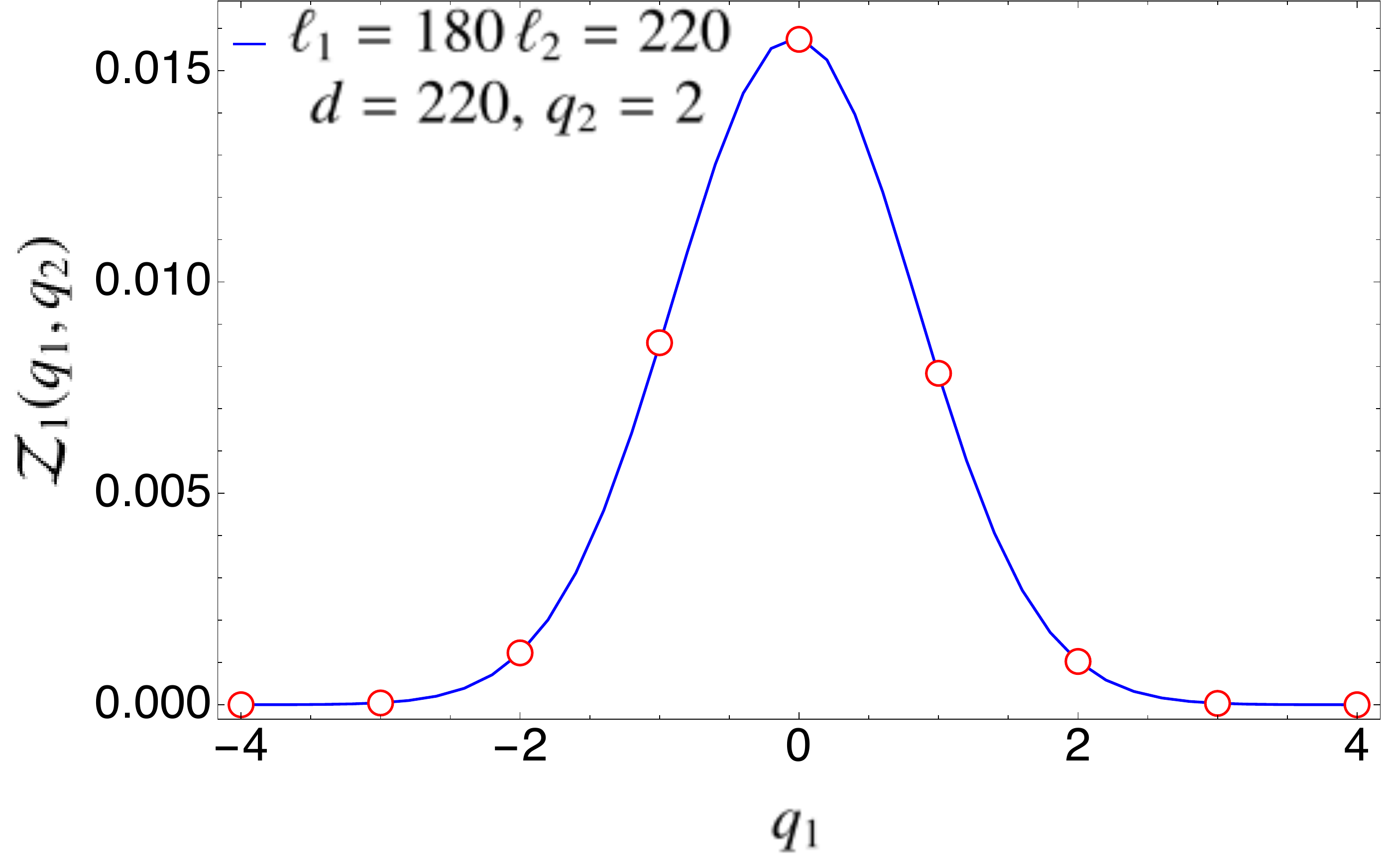}}
\subfigure
{\includegraphics[width=0.32\textwidth]{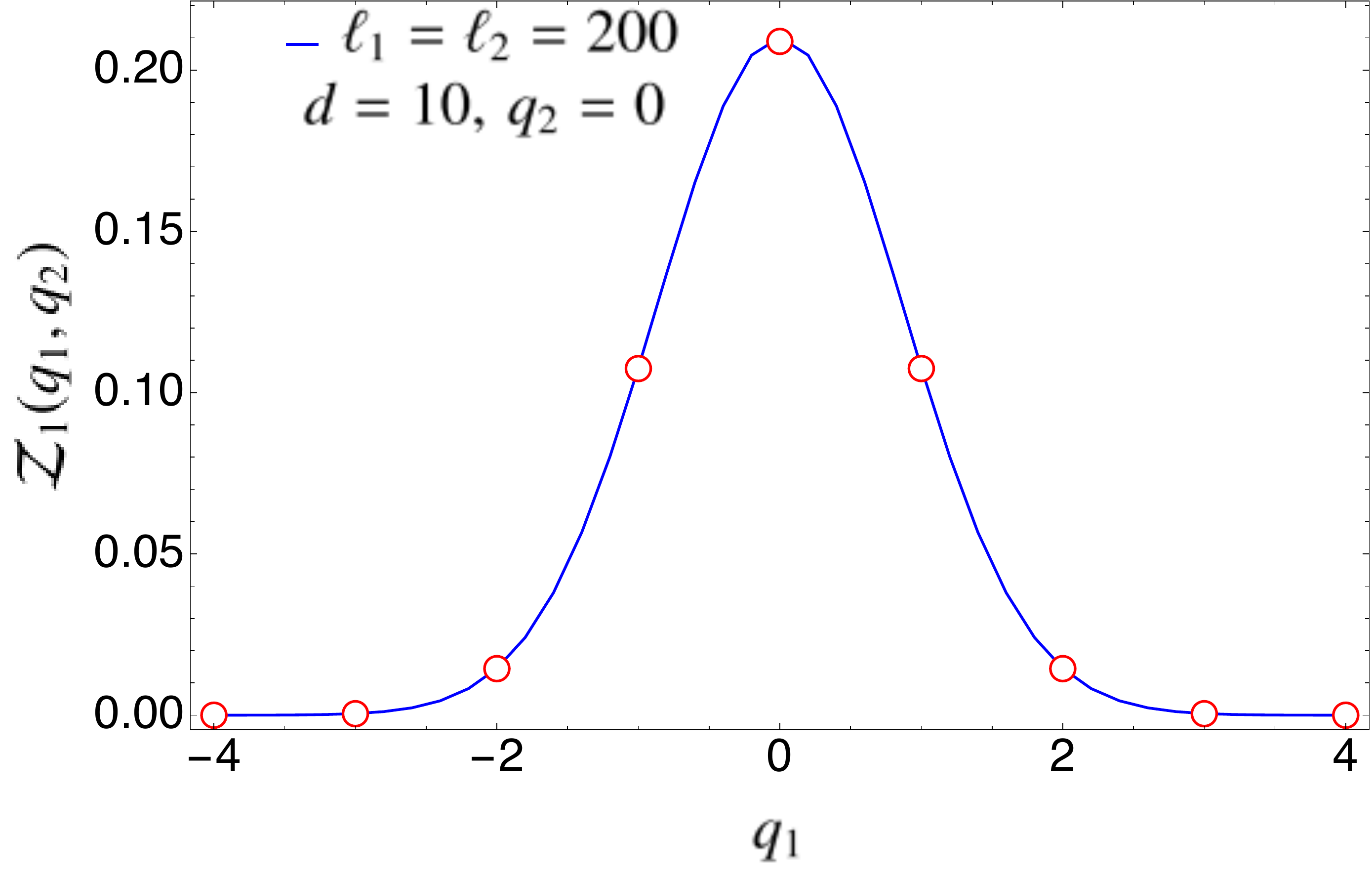}}
\subfigure
{\includegraphics[width=0.32\textwidth]{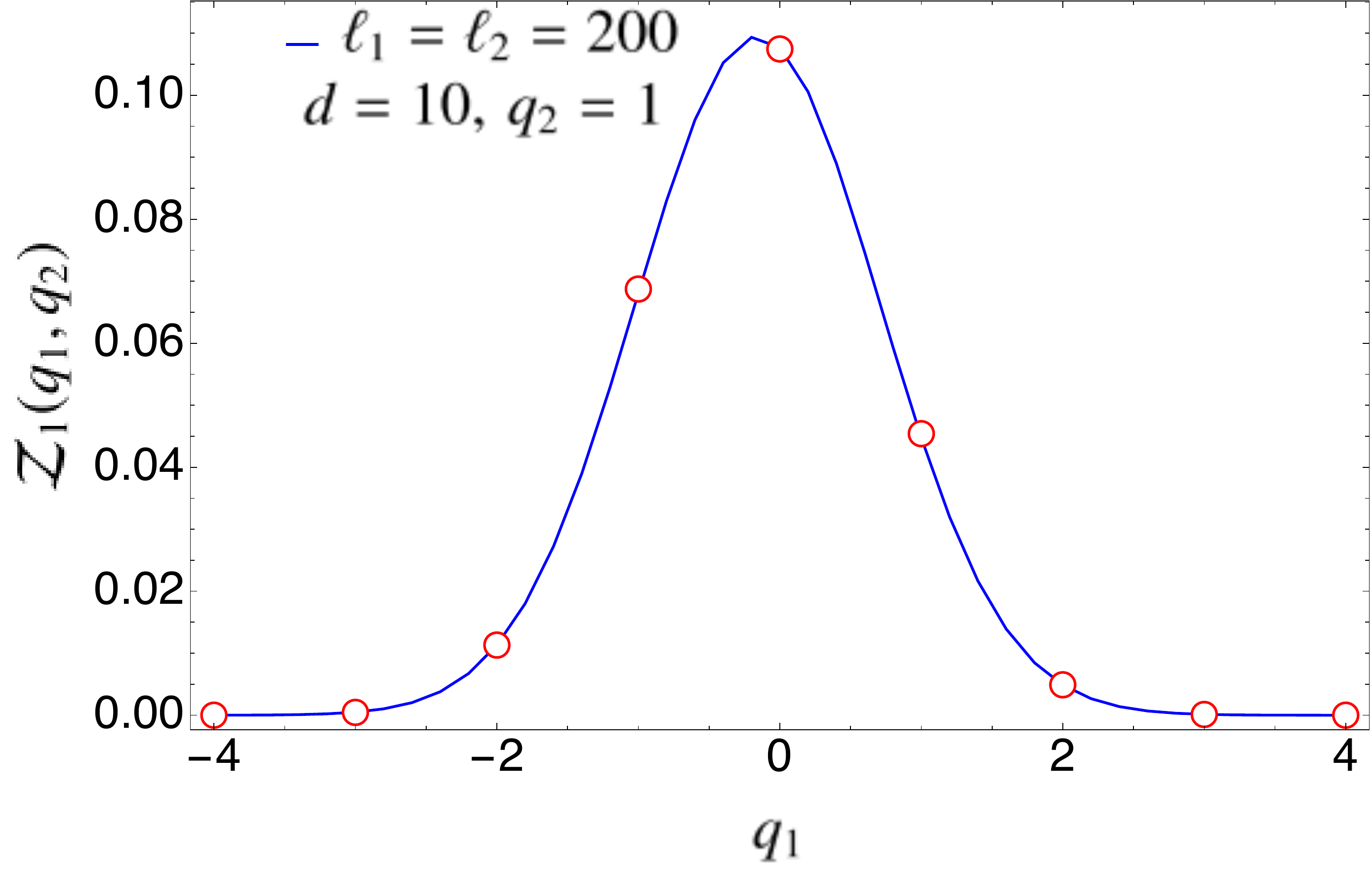}}
\subfigure
{\includegraphics[width=0.32\textwidth]{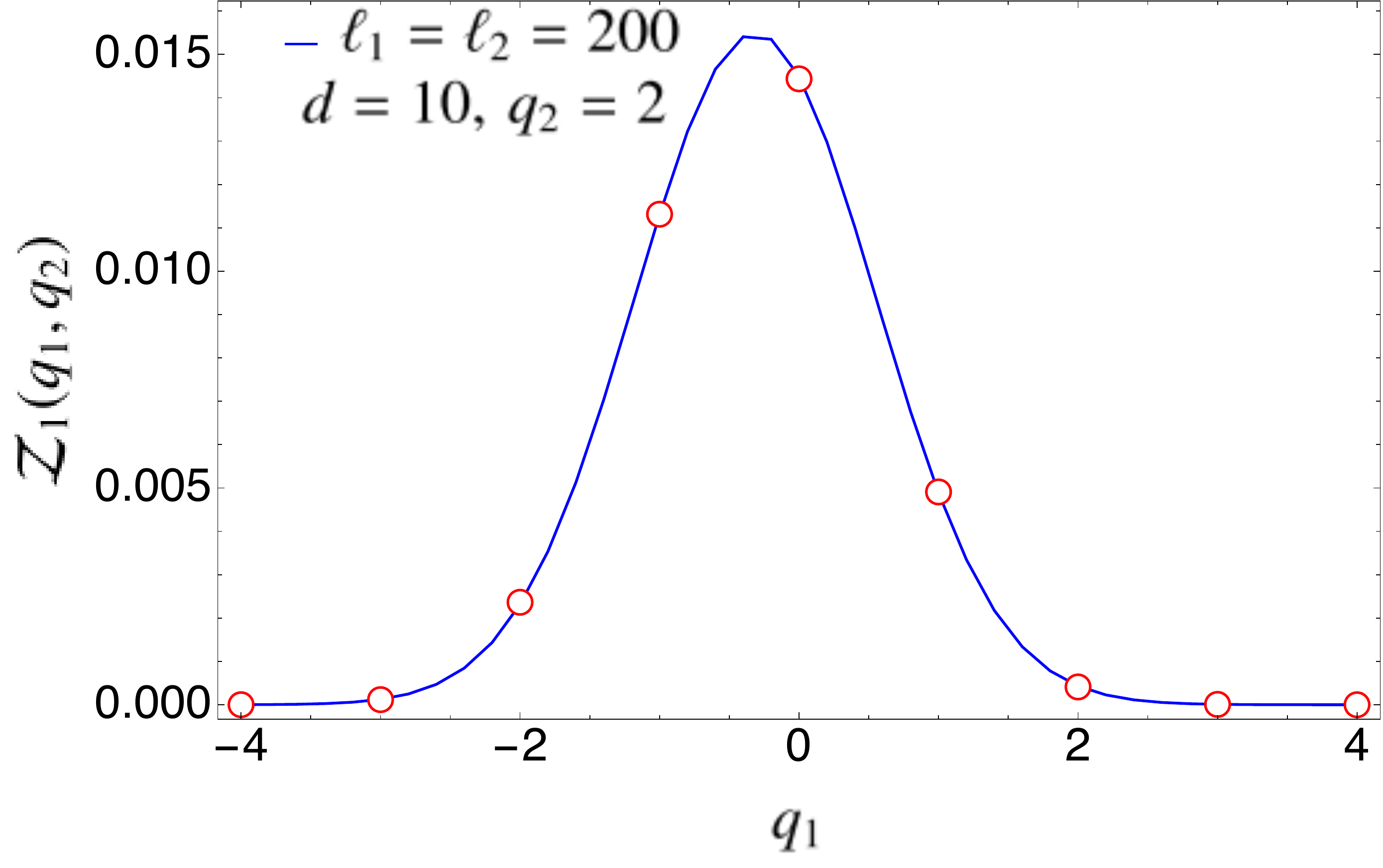}}
\caption{Probability $\mathcal{Z}_1(q_1,q_2)$ for the tight-binding model 
as a function of $q_1$ at fixed $q_2$ and for two disjoint intervals of lengths $\ell_1$, $\ell_2$, and separated by a distance $d$. 
The points are the exact numerical values calculated using the methods of Appendix~\ref{app:numericaltools}.
The solid line is the theoretical prediction in Eq.~\eqref{eq:Znq_alphabeta_disj} taking for the 
non-universal constants the corresponding values for the tight-binding model indicated in the main text. }
\label{fig:bn}
\end{figure}

\begin{figure}[t]
\centering
\subfigure
{\includegraphics[width=0.32\textwidth]{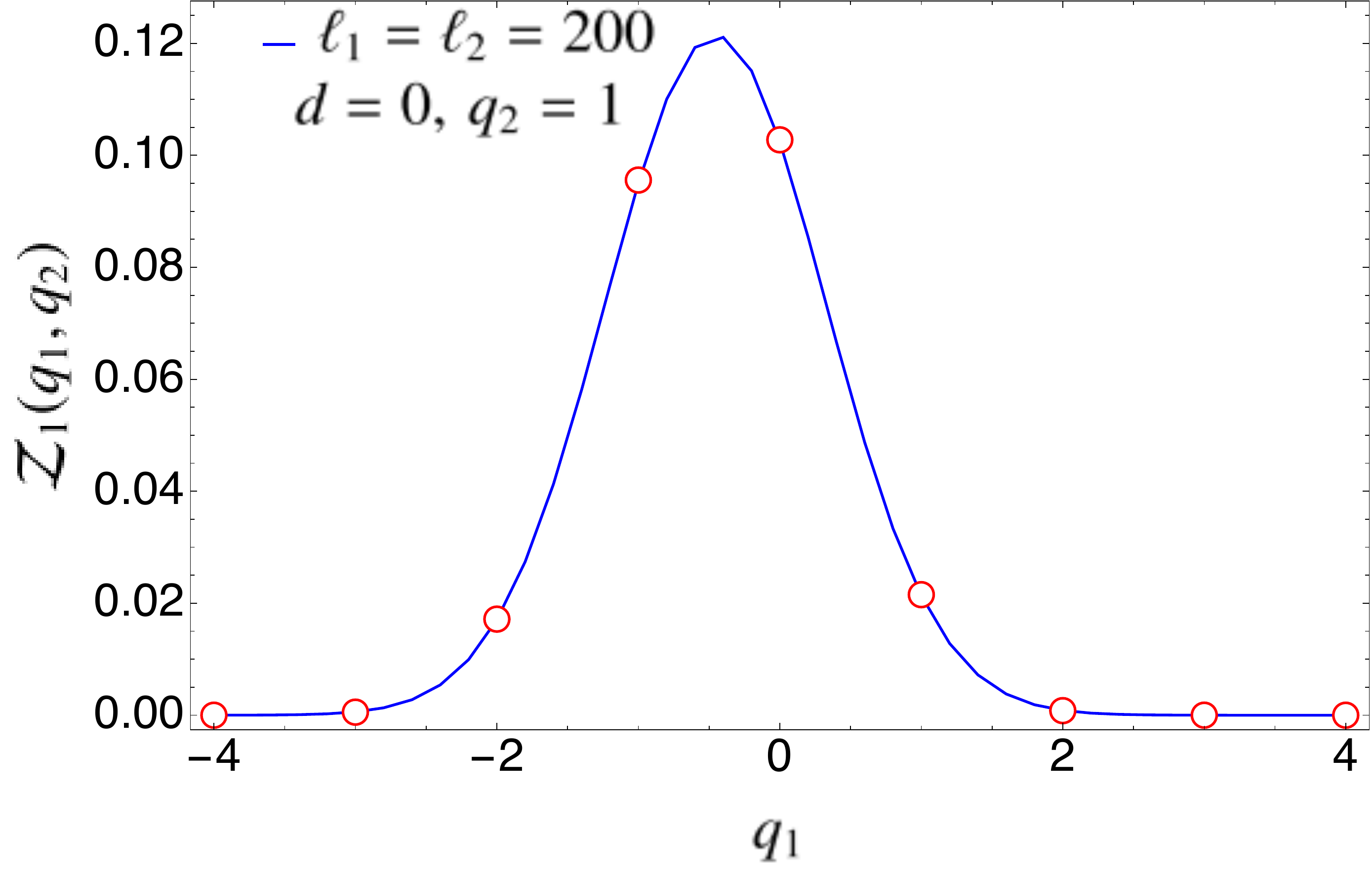}}
\subfigure
{\includegraphics[width=0.32\textwidth]{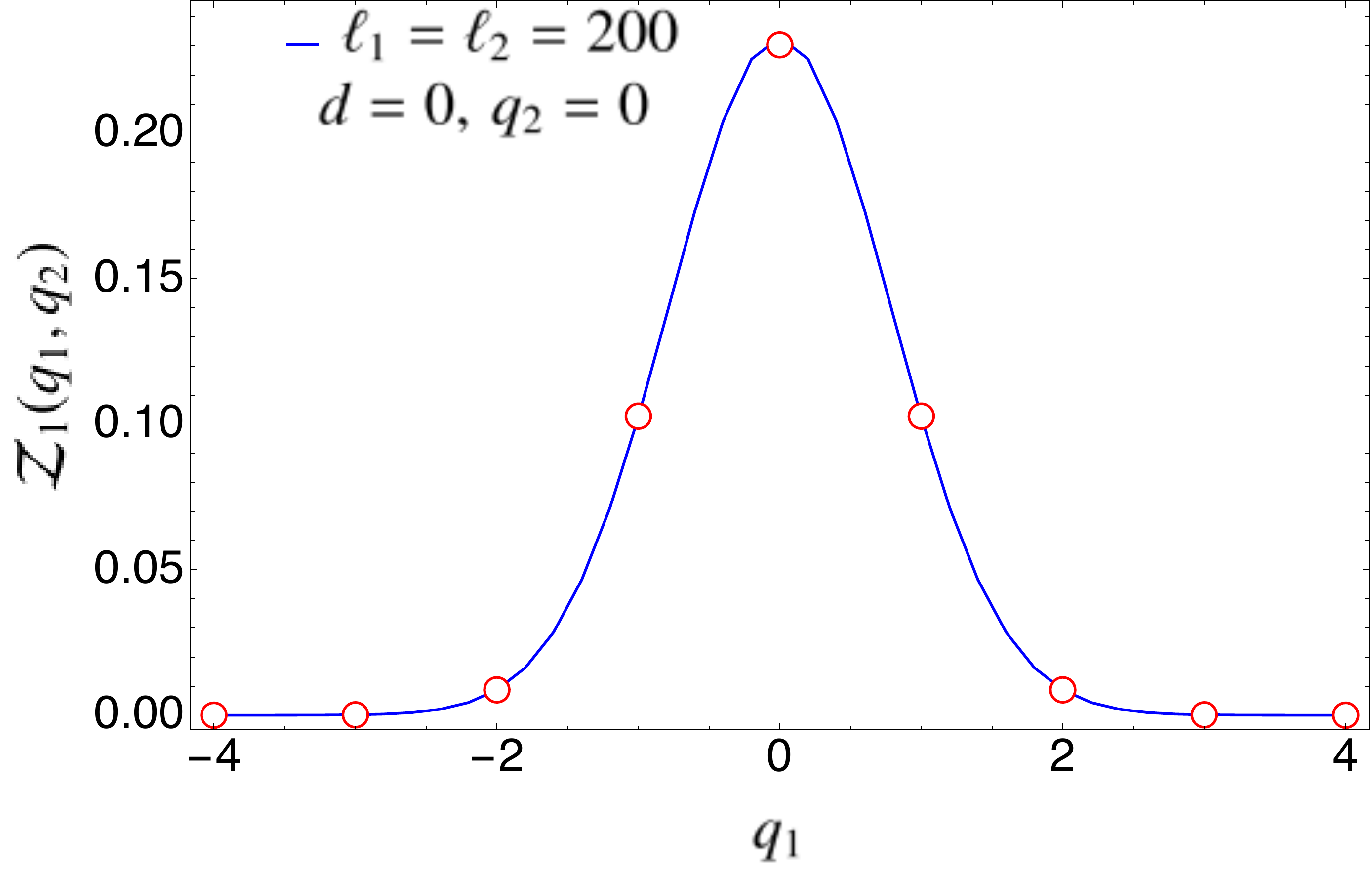}}
\subfigure
{\includegraphics[width=0.32\textwidth]{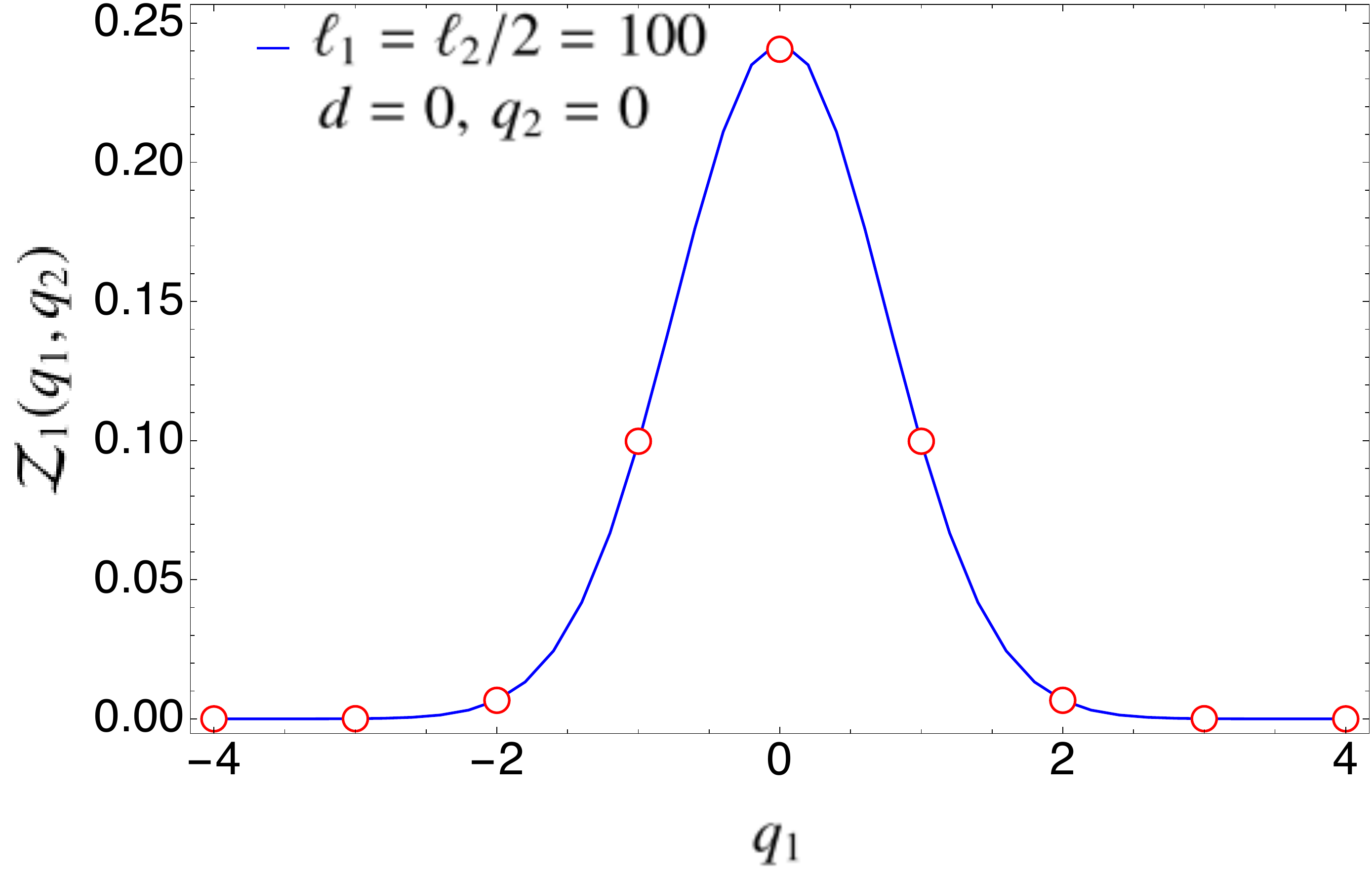}}
\caption{Probability $\mathcal{Z}_1(q_1,q_2)$ as a function of $q_1$ at fixed $q_2$ for two adjacent intervals of sizes $\ell_1$ and $\ell_2$. The solid blue line is the theoretical prediction in Eq.\,(\ref{eq:Znq_alphabeta_adj}). }
\label{fig:bn2}
\end{figure}

In Fig. \ref{fig:bn}, we compare the expression for
$\mathcal{Z}_1(q_1, q_2)$ found in Eq. \eqref{eq:Znq_alphabeta_disj}
for the case of disjoint intervals with the exact numerical results obtained for the tight-binding model using the methods of Appendix~\ref{app:numericaltools}.  The agreement is excellent. We remark that in Fig. \ref{fig:bn} there is no free parameter when matching the analytical
prediction with the numerical data since we know the expression of the non-universal constants for this particular system. In Fig.~\ref{fig:bn2}, we have repeated the same analysis in the case of adjacent intervals 
($d=0$), checking the validity of Eq.~\eqref{eq:Znq_alphabeta_adj}.

\subsection{Symmetry-resolved mutual information}
\begin{figure}[t!]
\centering
{\includegraphics[width=0.48\textwidth]{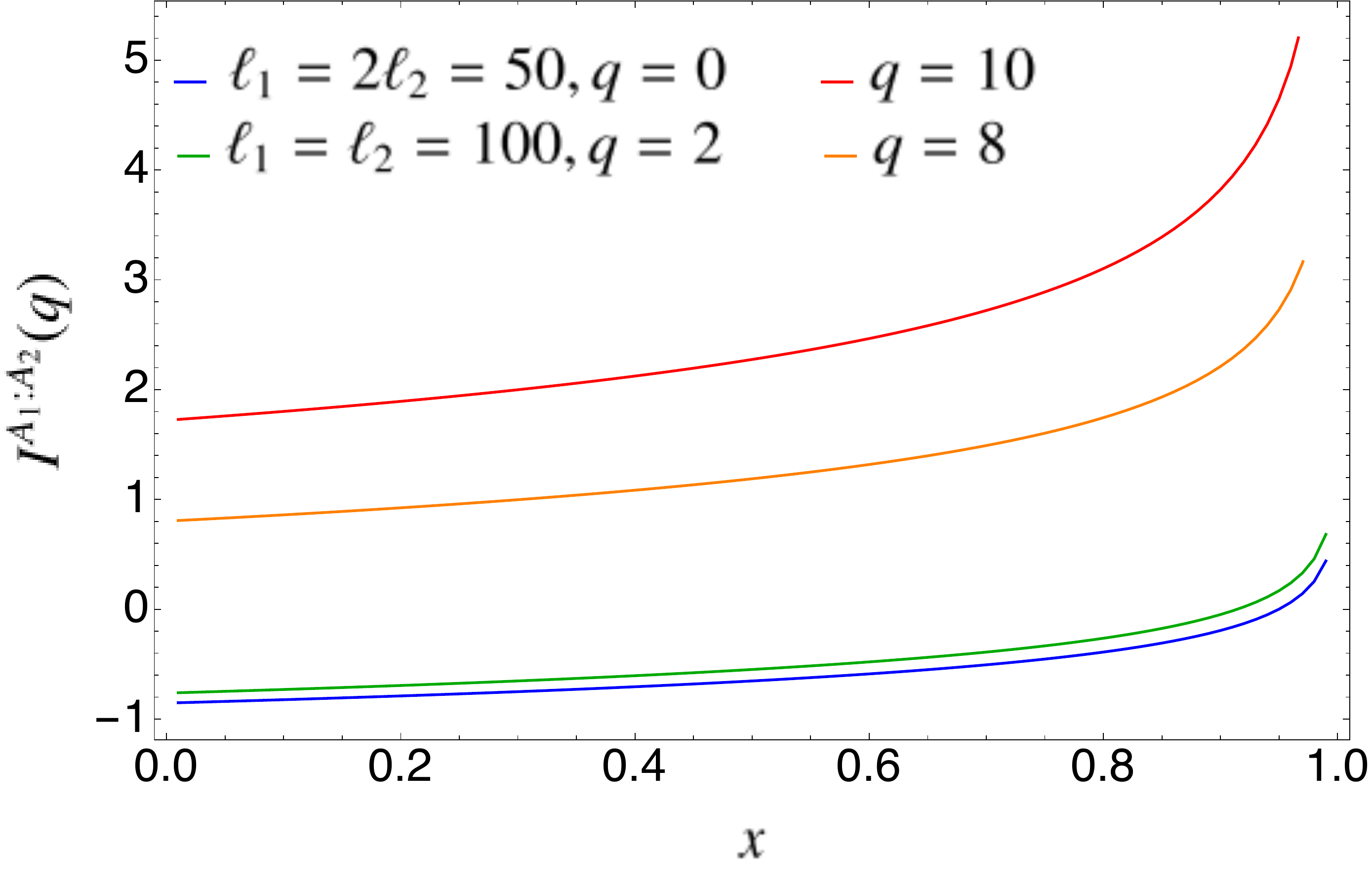}}
\subfigure
{\includegraphics[width=0.48\textwidth]{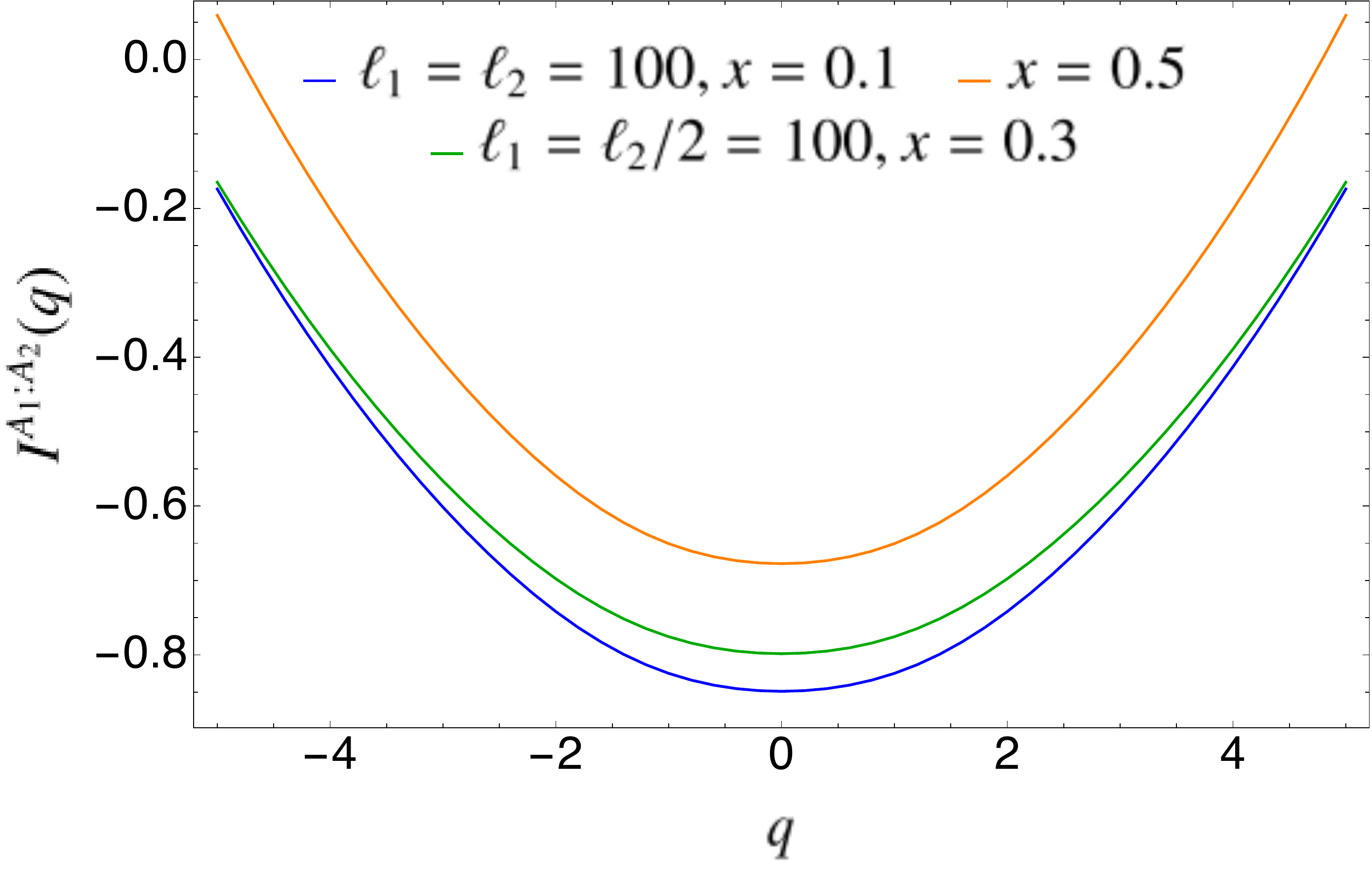}}
\caption{Symmetry-resolved mutual information of Eq. \eqref{eq:SRMIDirac} in the tight-binding model. We plot our analytical prediction for different combinations of $ \ell_1, \ell_2$ as a function of the cross-ratio $x$ (left panel) and $q$ (right panel).}
\label{fig:mutual}
\end{figure}

We compute now the symmetry-resolved mutual information defined in Eq.~(\ref{eq:SRMI}). 
We need the probability $p(q_1,q-q_1)$ derived in Eq.~\eqref{eq:cond_prob_final} as well 
as the symmetry-resolved entropies for $A$ and its parts $A_1$ and $A_2$ separately.
For the entropies of $A_1$ and $A_2$, we can use the results for a single interval obtained
in Ref.~\cite{riccarda} while, for the full subsystem $A$, it can be derived from the Fourier transform of the charged moments determined in Eq.~\eqref{eq:fourier_tranf_non_charg_mom} by applying Eq.~\eqref{eq:SRREfromqmoments}. The three symmetry-resolved entropies can eventually
be written in the form
\begin{equation}
\label{eq:SREEsingleordisjoint}
S_1^{X}(q)=
S_{1}^{X}
-\frac{1}{2}\ln\left(\frac{2K}{\pi}\ln\delta^\sigma\Lambda\right)
-\frac{1}{2}-\sigma \pi^4 \frac{\xi^2}{(K \ln(\lambda_1^\sigma\Lambda) )^2} 
+\sigma q^2\pi^4 \frac{\xi}{(K \ln(\lambda_1^\sigma\Lambda) )^2}. 
\end{equation}
where $S_1^X$ is the total entanglement entropy of subsystem $X$.
In this expression, when $X=A_p$, we have to take $\Lambda=\ell_p$ and $\sigma=1$ while, if $X=A_1\cup A_2$, then $\Lambda=\ell_1\ell_2(1-x)$ and $\sigma=2$. The auxiliary quantities $\delta$ and $\xi$ in Eq.~(\ref{eq:SREEsingleordisjoint}) 
are defined in terms of $\lambda_{n}$ as
\begin{equation}
\ln \delta=\ln \lambda_1+2\pi^2\xi, \qquad
   \xi=- \frac{1}{2\pi^2}\partial_n(\ln \lambda_n)|_{n=1}
\end{equation}
For the tight-binding model, we know the explicit value of these non-universal constants,
\begin{equation}
\ln\delta=2\pi^2 \gamma'_2(1)+\ln 2,
\qquad
\xi=\gamma_2(1)+\gamma'_2(1).
\end{equation}
From Eqs.~\eqref{eq:cond_prob_final} and 
\eqref{eq:SREEsingleordisjoint}, we can now obtain an explicit expression 
for the symmetry-resolved mutual information.
Since the conditional probability $p(q_1,q-q_1)$ satisfies Eq. 
(\ref{eq:normalisation}), we have
\begin{eqnarray}
\label{eq:SRMIDirac}
I^{A_1: A_2}(q)&=&I^{A_1:A_2} 
-\frac{1}{2}\ln\left[
\frac{2K}{\pi}\frac{\ln(\tilde{\ell}_1^{\delta}) \ln(\tilde{\ell}_2^{\delta})}
{\ln\left(\tilde{\ell}_1^{\delta}\tilde{\ell}_2^\delta(1-x)\right)}\right]
-\frac{1}{2}
\textcolor{blue}{-}
2 q^2\pi^4 \frac{\xi}{K^2 \ln^2\left(\tilde{\ell}_1
\tilde{\ell}_2(1-x)\right)}
\nonumber
\\
&&-  \pi^4 \frac{\xi^2}{K ^2}
\left(\frac{1}{\ln\tilde{\ell}_1} +\frac{1}{\ln\tilde{\ell}_2} 
-\frac{1}{\ln(\tilde{\ell}_1\tilde{\ell}_2(1-x))} \right)\nonumber\\
&&+\frac{\pi^4}{K^2} 
\xi
\int_{-\infty}^\infty p(q_1,q-q_1)
\left[
\frac{q_1^2}{\ln^2\tilde{\ell}_1}+\frac{(q-q_1)^2}{ \ln^2\tilde{\ell}_2}
\right] {\rm d}q_1,
\end{eqnarray}
where
$I^{A_1:A_2}$ is the total mutual information of Eq.~\eqref{eq:defMI} and we have introduced the rescaled 
subsystem length $\tilde{\ell}_p^{\delta}=\delta\ell_p$.
We plot this function in Fig. \ref{fig:mutual}. As we anticipated, the symmetry-resolved mutual information is not a good measure of the total correlations between $A_1$ and $A_2$ in each symmetry sector since it can assume negative values. 
Finally, we can also derive the number mutual information
defined in Eq.~(\ref{eq:totalMI}). Applying in that formula the result for $I^{A_1:A_2}(q)$ obtained in Eq.~(\ref{eq:SRMIDirac}), we have
\begin{eqnarray}
\label{eq:numberMIDirac}
I^{A_1:A_2}_{\textrm{num}}&=& \frac{1}{2}\ln\left[
\frac{2K}{\pi}\frac{\ln(\tilde{\ell}_{1}^\delta) \ln(\tilde{\ell}_2^\delta)}
{\ln\left(\tilde{\ell}_1^\delta\tilde{\ell}_2^\delta(1-x)\right)}
\right]
+\frac{1}{2}
+
\frac{2\pi^2\xi}
{K \ln\left(\tilde{\ell}_1\tilde{\ell}_2(1-x)\right)}\nonumber
\\&&-  \pi^4 \frac{\xi^2}{K ^2}
\left(\frac{1}{\ln\tilde{\ell}_1} +\frac{1}{\ln\tilde{\ell}_2} -
\frac{1}{\ln(\tilde{\ell}_1\tilde{\ell}_2(1-x))} \right)\nonumber\\
&&-
 \pi^4 
\frac{\xi}{K^2}
\int_{-\infty}^\infty \mathcal{Z}_1^{A_1:A_2}(q_1,q-q_1)
\left[
\frac{q_1^2}{\ln^2\tilde{\ell}_1 }+\frac{(q-q_1)^2}{\ln^2\tilde{\ell}_2}
\right] {\rm d}q_1 {\rm d}q.
\nonumber
\\
\,
\end{eqnarray}
Since $q=q_1+q_2$ and 
\begin{equation}
\int_{-\infty}^\infty \mathcal{Z}_1^{A_1:A_2}(q_1,q_2)
q_p^2 {\rm d}q_1 {\rm d}q_2
=
\frac{K\ln \tilde{ \ell_p}}{\pi^2},\quad p=1, 2,
\end{equation}
then Eq. (\ref{eq:numberMIDirac}) becomes
\begin{multline}
I^{A_1: A_2}_{\textrm{num}}= \frac{1}{2}\ln\left[
\frac{2K}{\pi}\frac{\ln(\tilde{\ell}_1^\delta) \ln(\tilde{\ell}_2^\delta)}
{\ln(\tilde{\ell}_1^\delta\tilde{\ell}_2^\delta(1-x))}
\right]
+\frac{1}{2}
 -  \pi^4 \frac{\xi^2}{K ^2}
\left(\frac{1}{\ln\tilde{\ell}_1} +\frac{1}{\ln\tilde{\ell}_2} 
-\frac{1}{\ln(\tilde{\ell}_1\tilde{\ell}_2(1-x))} 
 \right)\\
 -
 \pi^2 
\frac{\xi}{K}\left(\frac{1}{\ln\tilde{\ell}_1} 
+ \frac{1}{\ln\tilde{\ell}_2}
-
 \frac{2}
 {\ln(\tilde{\ell}_1\tilde{\ell}_2(1-x))}\right).
\end{multline}
In the limit $\ell_1,\ell_2, d \to \infty$, this expression behaves as
\begin{equation}\label{eq:numasymp}
I^{A_1: A_2}_{{\rm num}}\sim \frac{1}{2}\ln\left[
\frac{2K}{\pi}\frac{\ln\ell_1 \ln\ell_2}{\ln\left(\ell_1\ell_2(1-x)\right)}
\right]
+\frac{1}{2}.
\end{equation} 
This result resembles the one for the number entropy of a single 
interval (see e.g. \cite{riccarda}), where a double logarithmic 
correction in the subsystem length also appears, even though, in our 
case, the dependence on the parameters $\ell_1,\ell_2,d$ is more 
involved. On the other hand, it is a simple function of the 
Luttinger parameter $K$ since, as we already pointed out, the only 
effect of $K$ in the Gaussian factor of the Fourier transform of the 
multi-charged moments is renormalising its variance.

\section{Conclusions}\label{sec:concl}

In this manuscript, we have computed the multi-charged moments
of two intervals in the ground state of the free massless Dirac 
field  and the massless compact boson, with arbitrary compactification radius.
Using the replica approach, the multi-charged moments are given by the partition
function of the theory on a higher genus Riemann surface with a different magnetic
flux inserted in each interval. We have carried out the analysis of such partition
function for the two CFTs under investigation in full generality, allowing  the 
background magnetic flux to generate a different twisted boundary condition at each end-point of the intervals. 
In the case of the Dirac field, we have adapted
the diagonalisation in the replica space method of Ref.~\cite{CFH}, to account for the 
different monodromy of the fields at each end-point. In the compact boson theory, 
we have chosen a geometric approach, and we have directly considered the four-point correlator on the 
Riemann surface of the vertex operators that implement the flux. It turns out that the 
known formulae for such correlator diverge in our case~\cite{Verlinde}. Once we properly regularised them, 
we have obtained a cumbersome expression for the multi-charged moments in terms of Riemann-Siegel Theta functions. 
Nevertheless, we have found several remarkable identities concerning the prime forms 
of the Riemann surface that allow to dramatically simplify 
the final result for the multi-charged moments. The factor due to the magnetic fluxes is eventually an
algebraic function of the lengths of the intervals and their separation---identical to the one 
obtained in the Dirac theory. In fact, for a certain value of the compactification radius,
the multi-charged moments of both theories are equal, generalising the known identity between 
their two-interval R\'enyi entropies~\cite{HLM13}.

Given the simple expression obtained for the multi-charged moments, we can easily calculate their 
Fourier transform, which has a Gaussian form. From it, we have finally derived formulae
for several interesting quantities such as the joint probability distribution of the charge for
simultaneous measures in the two intervals, the symmetry-resolved mutual information~\cite{pbc-21} and the number mutual information.

Let us conclude this manuscript discussing few outlooks. The multi-charged moments analysed here can 
be used to study the symmetry decomposition of the negativity in imbalance sectors \cite{goldstein1}. This is a measure 
of entanglement in mixed states which involves the partial transposition of the reduced density matrix. 
In the replica approach, this operation can be performed by properly fixing the different fluxes and exchanging 
the end-points of the transposed interval.  A further generalisation is to identify the holographic 
dual of the multi-charged moments, which would be the starting point to compute the symmetry-resolved 
mutual information in the AdS/CFT correspondence, as done for the entanglement entropy in Ref.~\cite{znm-20}.
Partition functions with twisted boundary conditions, as the ones considered here, have been also proposed 
as non-local order parameters to distinguish various topological phases of spin chains \cite{ssr1-17,ssr-16}. 
We think that our analysis for the multi-charged moments can be useful to make progresses also in that direction.
Finally, even though far beyond the scope of this manuscript, it would be interesting to obtain a rigorous proof 
of the identities for the prime forms that we have found and numerically checked here.

\section*{Acknowledgments}
We thank Benoit Estienne, Yacine Ikhlef and Tamara Grava for useful discussions. 
PC, FA and SM acknowledge support from ERC under Consolidator grant number 771536 (NEMO).

\appendix
\section*{Appendices}

\section{Numerical Methods}
\label{app:numericaltools}

For the numerical test of our field theory results, we consider the following lattice discretisation of the Dirac fermion, known as tight-binding model,
\begin{equation}\label{eq:hamiltonianF}
H=-\frac{1}{2}\sum_{j=-\infty}^{\infty}(c^{\dagger}_{j+1}c_j+\mathrm{h.c.}),
\end{equation}
where  $c_j^\dagger$ and $c_j$ are fermionic creation and annihilation operators that 
satisfy the anti-commutation relations $\{ c_j,c^{\dagger}_k \}=\delta_{jk}$. 
In terms of them, the charge operator reads
\begin{equation}
Q=\sum_{j=-\infty}^\infty \left(c^{\dagger}_j c_j-\frac{1}{2}\right).
\end{equation}
The two-point correlation functions in the ground state of Eq.~\eqref{eq:hamiltonianF} are of the form 
\begin{equation}
\braket{c^{\dagger}_jc_{k}}=\frac{\sin(\pi(j-k))}{2\pi(j-k)},
\end{equation}
and, due to the particle number conservation, $\langle c_j c_k\rangle=0$. As well-known~\cite{p-03,pe-09}, the moments ${\rm Tr}[\rho_A^n]$ can be calculated from the restriction of the two-point correlation matrix to the subsystem $A$, that
is $(C_A)_{j, k}=\langle c_j^\dagger c_k\rangle$, with $j,k\in A$. The charged moments $Z^{A=A_1 \cup A_2}_{n}(\alpha)$ can also be
easily computed numerically in terms of the matrix $C_A$ using the 
formula~\cite{goldstein, riccarda}
\begin{equation}\label{eq:Zalphanumerics}
Z_n^{A=A_1 \cup A_2}(\alpha)=\displaystyle \prod_{j=1}^{\ell_1+\ell_2} [(\varepsilon_{j})^n e^{i\alpha/2}+(1-\varepsilon_{j} )^n e^{-i\alpha/2}],
\end{equation}
where $\varepsilon_j$ are the eigenvalues of $C_A$ and $\ell_p$ is the number of sites in the interval $A_p$.
In the case of the multi-charged moments $Z^{A_1:A_2}(\alpha,\beta)$
defined in Eq. \eqref{eq:Charged_alphabeta},  the method used to 
compute $Z_n^{A}(\alpha)$ can not be applied since $\rho_{A}$ does not commute with the charges $Q_{A_1}$ and 
$Q_{A_2}$ of the two parts of $A$. Following Ref.~\cite{pbc-21} (which was based on \cite{gec-18}), we rewrite Eq. \eqref{eq:Charged_alphabeta} as
\begin{equation}
    Z^{A_1:A_2}_1(\alpha,\beta)=\tilde{Z}_A\mathrm{Tr}_A(\rho_A\tilde{\rho}_A),
\end{equation}
where
\begin{equation}
   \tilde{\rho}_A=\frac{1}{\tilde{Z}_A}e^{i\alpha Q_{A_1}+i\beta Q_{A_2}}, \, \quad \tilde{Z}_A= \mathrm{Tr}_A(e^{i\alpha Q_{A_1}+i\beta Q_{A_2}} ).
\end{equation}
Although $ \tilde{\rho}_A$ is not a density matrix, it is a Gaussian operator with an associated two-point correlation matrix, $\tilde{C}_A$, given by
\begin{equation}
    (\tilde{C}_A)_{kj}=\delta_{kj}\begin{cases}
    \frac{e^{i\alpha}}{1+e^{i\alpha}} \quad j \in A_1, \\
    \frac{e^{i\beta}}{1+e^{i\beta}}\quad j \in A_2.
    \end{cases}
\end{equation}
Applying the rules for the composition of Gaussian operators~\cite{fc-10} and introducing $W=2C_A-\mathbb{I}$, we get~\cite{pbc-21}
\begin{equation}
   Z^{A_1:A_2}_1(\alpha,\beta)= (e^{-i\alpha/2} + e^{i\alpha/2})^{\ell_1}(e^{-i\beta/2} + e^{i\beta/2})^{\ell_2}\mathrm{det}\left(\frac{\mathbbm{1}_{\ell_1+\ell_2}+W_{\alpha \beta}}{2}\right),
\end{equation}
where 
\begin{equation}
    W_{\alpha \beta}=\begin{pmatrix}
    W_{11} & W_{12}\\
    W_{21} & W_{22}\\
    \end{pmatrix}\begin{pmatrix}
    \frac{e^{i\alpha}-1}{e^{i\alpha}+1} \mathbbm{1}_{\ell_1}& 0\\
    0 & \frac{e^{i\beta}-1}{e^{i\beta}+1} \mathbbm{1}_{\ell_2} \\
    \end{pmatrix},
\end{equation}
and the notation $W_{pp'}$, $p,p'=1,2$, refers to correlations between sites in $A_p$ and $A_{p'}$. This result allows to exactly compute the multi-charged moments in the tight-binding model for different values of $\alpha$ and $\beta$, as showed in Fig. \ref{fig:Chargedmomentsdnon0}.
The Fourier transform of $Z_1^{A_1:A_2}(\alpha,\beta)$ gives the quantities $\mathcal{Z}_1^{A_1:A_2}(q_1,q_2)$ analysed in Fig. \ref{fig:bn} and Fig. \ref{fig:bn2}.

\section{Derivation of the normalised holomorphic differentials}\label{app:riemann}
\begin{figure}[t!]
\centering
\includegraphics[width=0.48\textwidth]{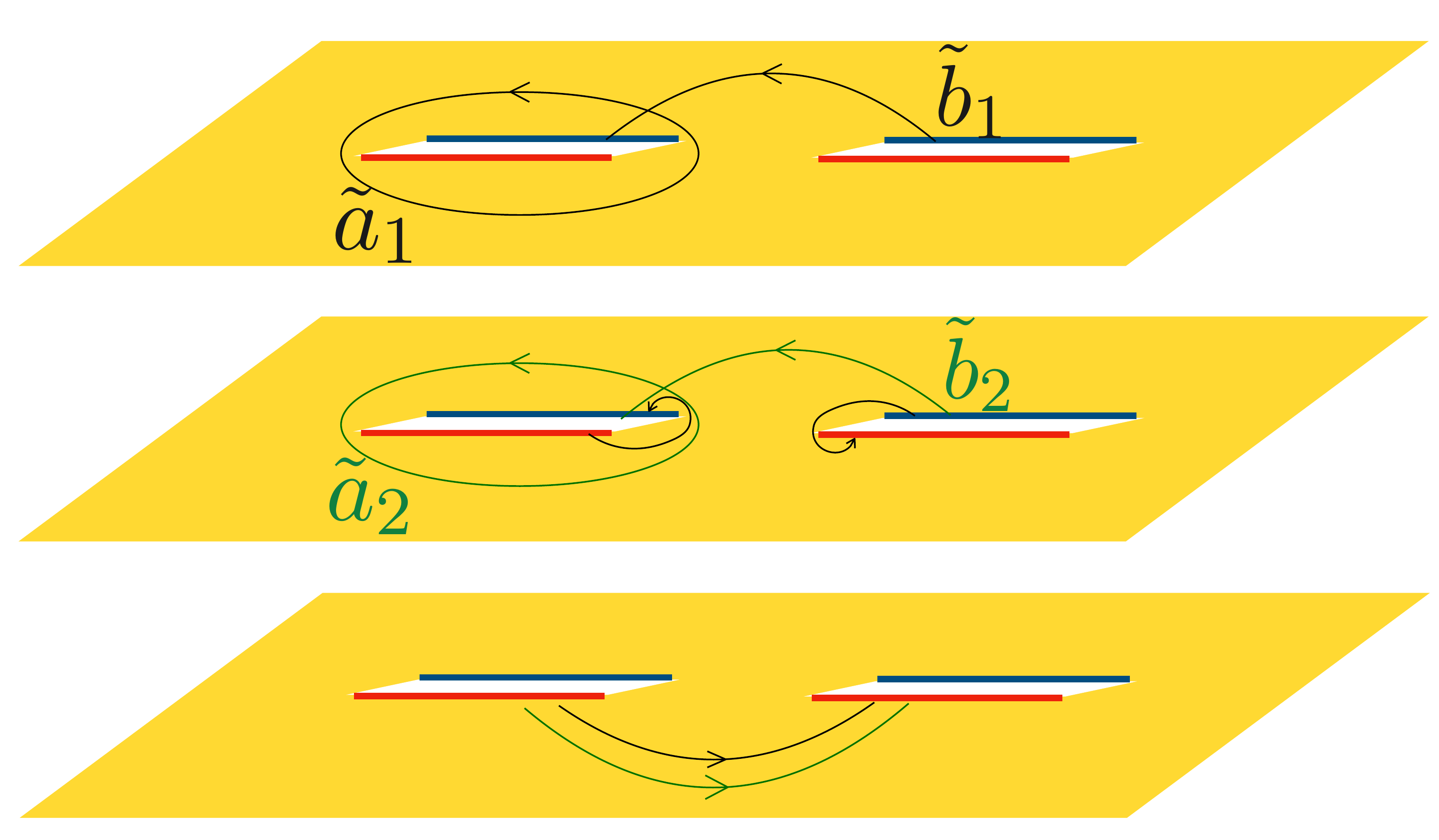}
\caption{Auxiliary homology basis of the surface $\Sigma_n(x)$ for $n=3$.}
\label{fig:rsurface}
\end{figure}

In this Appendix, we present the detailed derivation of the expression
given in Eq.~\eqref{eq:norm_diff} for the normalised holomorphic 
differentials $\nu_r(z)$ of the Riemann surface $\Sigma_n(x)$. 
Recall that, according to Eq.~\eqref{eq:norm_cond_nu}, they are normalised with respect to the homology basis $\{a_r,b_r\}$. In order
to define this basis, it is convenient to introduce an
auxiliary basis of non-contractible cycles, which we call 
$\tilde{a}_r$ and $\tilde{b}_r$. The cycle $\tilde{a}_r$ encloses 
anticlockwise the branch cut $(u_1, v_1)$ in the $r$-sheet of the 
surface. The dual cycle $\tilde{b}_r$ connects anticlockwise the $r$
and $n$ sheets through the branch cut $(u_1, v_1)$ and then it goes
back to the $r$ sheet through the branch cut $(u_2, v_2)$. 
In Fig.~\ref{fig:rsurface}, we draw the auxiliary homology
basis in the case $n=3$. The cycles $a_r$ and $b_r$ are defined 
in terms of the auxiliary ones by
\begin{equation}\label{eq:hom_change_basis}
 a_r=\sum_{k=1}^r \tilde{a}_k, \quad b_r=\tilde{b}_r-\tilde{b}_{r+1}, 
\end{equation}
with $r=1, \dots, n-1$ and $\tilde{b}_{n}=0$.

We can now obtain $\nu_r(z)$ by following the usual procedure considered in the literature, see e.g.~\cite{Enolskii, Enolskii2}. 
We first take the basis of canonical holomorphic differentials of the surface $\Sigma_n(x)$,
\begin{equation}
 \mu_r(z)=(z(z-1))^{-r/n}(z-x)^{r/n-1}, \quad r=1, \dots, n-1.
\end{equation}
If we denote by $\mathcal{A}$ and $\mathcal{B}$ the $(n-1)\times (n-1)$ 
matrices with entries
\begin{equation}\label{eq:matrices_calA_calB}
 \mathcal{A}_{r, s}=\oint_{a_r}{\rm d}z\mu_s(z), \quad 
 \mathcal{B}_{r, s}=\oint_{b_r}{\rm d}z\mu_s(z),
\end{equation}
then the normalised holomorphic differentials $\nu_r(z)$ can be 
calculated from $\mu_r(z)$ such that
\begin{equation}\label{eq:rel_nu_mu}
 \nu_r(z)=\sum_{l=1}^{n-1} \mu_l(z) \mathcal{A}^{-1}_{l, r},
\end{equation}
where $\mathcal{A}^{-1}_{l, r}$ are the entries of the 
inverse of the matrix $\mathcal{A}$. Furthermore, by combining 
Eq.~\eqref{eq:def_matrix_periods} with Eqs.~\eqref{eq:matrices_calA_calB} 
and \eqref{eq:rel_nu_mu}, the matrix of periods $\Gamma(x)$ 
in Eq.~\eqref{eq:matrix_periods} can be obtained from 
\begin{equation}\label{eq:matrix_periods_AB}
 \Gamma(x)=\mathcal{B}\mathcal{A}^{-1}.
\end{equation}

To compute the matrices $\mathcal{A}$ and $\mathcal{B}$, 
it is useful to consider the auxiliary homology basis 
$\{\tilde{a}_r,\tilde{b}_r\}$. 
The advantage of taking this basis is that it is easier to calculate 
the contour integrals of $\mu_r(z)$ along the cycles $\tilde{a}_r$ and $\tilde{b}_r$
than along $a_r$ and $b_r$. In fact, let us denote by $\tilde{\mathcal{A}}$ and $\tilde{\mathcal{B}}$ 
the $(n-1)\times (n-1)$ matrices analogous to $\mathcal{A}$ and $\mathcal{B}$
but integrating now along the auxiliary cycles $\tilde{a}_r$ and $\tilde{b}_r$
respectively,
\begin{equation}
 \tilde{\mathcal{A}}_{r, s}=\oint_{\tilde{a}_r}{\rm d}z \mu_s(z), \quad 
 \tilde{\mathcal{B}}_{r, s}=\oint_{\tilde{b}_r}{\rm d}z \mu_s(z).
\end{equation}
Then, by taking the correct branches for $\mu_r(z)$, we find that 
\begin{eqnarray}
 \tilde{\mathcal{A}}_{r, s}&=& e^{\frac{2\pi i(r-1)s}{n}}\left(e^{-\frac{2\pi i s}{n}}-1\right)\int_0^x {\rm d} z \mu_s(z)\nonumber\\
 &=& 2\pi i e^{\frac{\pi i(2r-3)s}{n}} I_{s/n}(x),
\end{eqnarray}
and
\begin{eqnarray}
 \tilde{\mathcal{B}}_{r, s}&=&e^{\frac{2\pi i(r-1)s}{n}}\left(e^{-\frac{2\pi i rs}{n}}-1\right)\int_x^1 {\rm d}z \mu_s(z)\nonumber\\
 &=&-2\pi i e^{\frac{i\pi(r-3)s}{n}}\frac{\sin\left(\frac{\pi rs}{n}\right)}{\sin\left(\frac{\pi s}{n}\right)}
 I_{s/n}(1-x),
\end{eqnarray}
where we have employed the identities
\begin{equation}\label{eq:int_identities}
 \int_0^x {\rm d}z \mu_s(z)=-\frac{\pi}{\sin\left(\frac{\pi s}{n}\right)}I_{s/n}(x),\quad 
 \int_x^1{\rm d}z \mu_s(z)=\frac{\pi e^{-\frac{i\pi s}{n}}}{\sin\left(\frac{\pi s}{n}\right)}I_{s/n}(1-x).
\end{equation}

If we apply now the relation of Eq.~\eqref{eq:hom_change_basis}
between both homology basis, then we directly obtain the matrices
$\mathcal{A}$,
\begin{eqnarray}\label{eq:matrix_A}
 \mathcal{A}_{r, s}=\sum_{k=1}^{r} \tilde{\mathcal{A}}_{k, s}
 =2\pi i e^{\frac{i\pi(r-2)s}{n}}
 \frac{\sin\left(\frac{\pi r s}{n}\right)}{\sin\left(\frac{\pi s}{n}\right)}
 I_{s/n}(x),
\end{eqnarray}
and $\mathcal{B}$
\begin{eqnarray}\label{eq:matrix_B}
 \mathcal{B}_{r, s}&=&\tilde{\mathcal{B}}_{r, s}-\tilde{\mathcal{B}}_{r+1, s}\nonumber\\
 &=&2\pi i e^{\frac{i\pi(r-2)s}{n}}\left[\sin\left(\frac{\pi(r+1)s}{n}\right)
 -e^{-\frac{i\pi s}{n}}\sin\left(\frac{\pi rs}{n}\right)\right]\frac{I_{s/n}(1-x)}
 {\sin\left(\frac{\pi s}{n}\right)}.
\end{eqnarray}

The entries of the inverse of $\mathcal{A}$ are 
\begin{equation}\label{eq:inv_cal_A}
 \mathcal{A}_{r,s}^{-1}=\frac{e^{-\frac{2\pi i(s-1)r}{n}}\sin\left(\frac{\pi r}{n}\right)}{\pi n I_{r/n}(x)}.
\end{equation}
Using Eqs.~\eqref{eq:rel_nu_mu} and \eqref{eq:inv_cal_A}, we finally arrive at the expression written 
in Eq.~\eqref{eq:norm_diff} for the normalised holomorphic differentials $\nu_r(z)$. The matrix of periods in 
Eq.~\eqref{eq:matrix_periods} can be directly obtained by plugging Eqs.~\eqref{eq:matrix_B} and 
\eqref{eq:inv_cal_A} into Eq.~\eqref{eq:matrix_periods_AB}.

\section{Prime form identities}\label{app:prime_forms}

The identities for the regularised prime forms $E^{(*)}(z, z')$ of
Eqs.\eqref{eq:prime_form_id_1}-\eqref{eq:prime_form_id_4} can be easily proved when $n=2$ (genus one). In that 
case, the matrix of periods of Eq.~\eqref{eq:matrix_periods} is a scalar $\Gamma(x)={\rm i}\beta_{1/2}(x)$ and 
the Theta function $\Theta_{\frac{1}{2}}$ that appears in Eq.~\eqref{eq:reg_prime_form} reduces to a Jacobi theta
function, $\Theta_{\frac{1}{2}}(\boldsymbol{u}|\Gamma(x))=-\vartheta_1(\boldsymbol{u}|\Gamma(x))$. The images under the Abel-Jacobi map in Eqs.~\eqref{eq:abel_map_2}-\eqref{eq:abel_map_4} are 
in this case 
\begin{equation}
    \boldsymbol{w}(x)=\frac{1}{2},\quad 
    \boldsymbol{w}(1)=\frac{1}{2}+\frac{\Gamma(x)}{2},\quad 
    \boldsymbol{w}(\infty)=\frac{\Gamma(x)}{2}.
\end{equation}
Here we illustrate the proof of Eq.~\eqref{eq:prime_form_id_2}. The rest of
identities can be checked in a similar way by applying the different
relations in Sec. 20.2 of Ref.~\cite{nist}. If $n=2$,  Eq.~\eqref{eq:reg_prime_form} takes the following form for $z=x$, $z'=1$,
\begin{equation}
    |E^{(*)}(x, 1)|=2\pi x^{1/4}(1-x)^{1/2} I_{1/2}(x)
    \left|\frac{\vartheta_1(\Gamma(x)/2|\Gamma(x))}
    {\partial_u\vartheta_1(0|\Gamma(x))}\right|,
\end{equation}
where $I_p(x)$ is defined below Eq.~(\ref{eq:beta_def}).
If we apply the half-period translation $\vartheta_1(\Gamma/2|\Gamma)=i M_2(x)^{-1}\vartheta_4(0|\Gamma)$, the equality 
\begin{equation}
\partial_u\vartheta_1(0|\Gamma)=\pi \vartheta_2(0|\Gamma)\vartheta_3(0|\Gamma)\vartheta_4(0|\Gamma),
\end{equation}
and the relation between the hypergeometric function and the 
Jacobi theta function $I_{1/2}(x)=\vartheta_3(0|\Gamma)^2$,
we find
\begin{equation}
    |E^{(*)}(x, 1)|=\frac{2 x^{1/4}(1-x)^{1/2}}{M_2(x)}\frac{\vartheta_3(0|\Gamma)}{\vartheta_2(0|\Gamma)}.
\end{equation}
Finally, employing the well-known equality
\begin{equation}
    x^{1/4}=\frac{\vartheta_2(0|\Gamma)}{\vartheta_3(0|\Gamma)},
\end{equation}
we obtain Eq.~\eqref{eq:prime_form_id_2}.

\begin{figure}[htbp!]
\centering
\subfigure
{\includegraphics[width=0.48\textwidth]{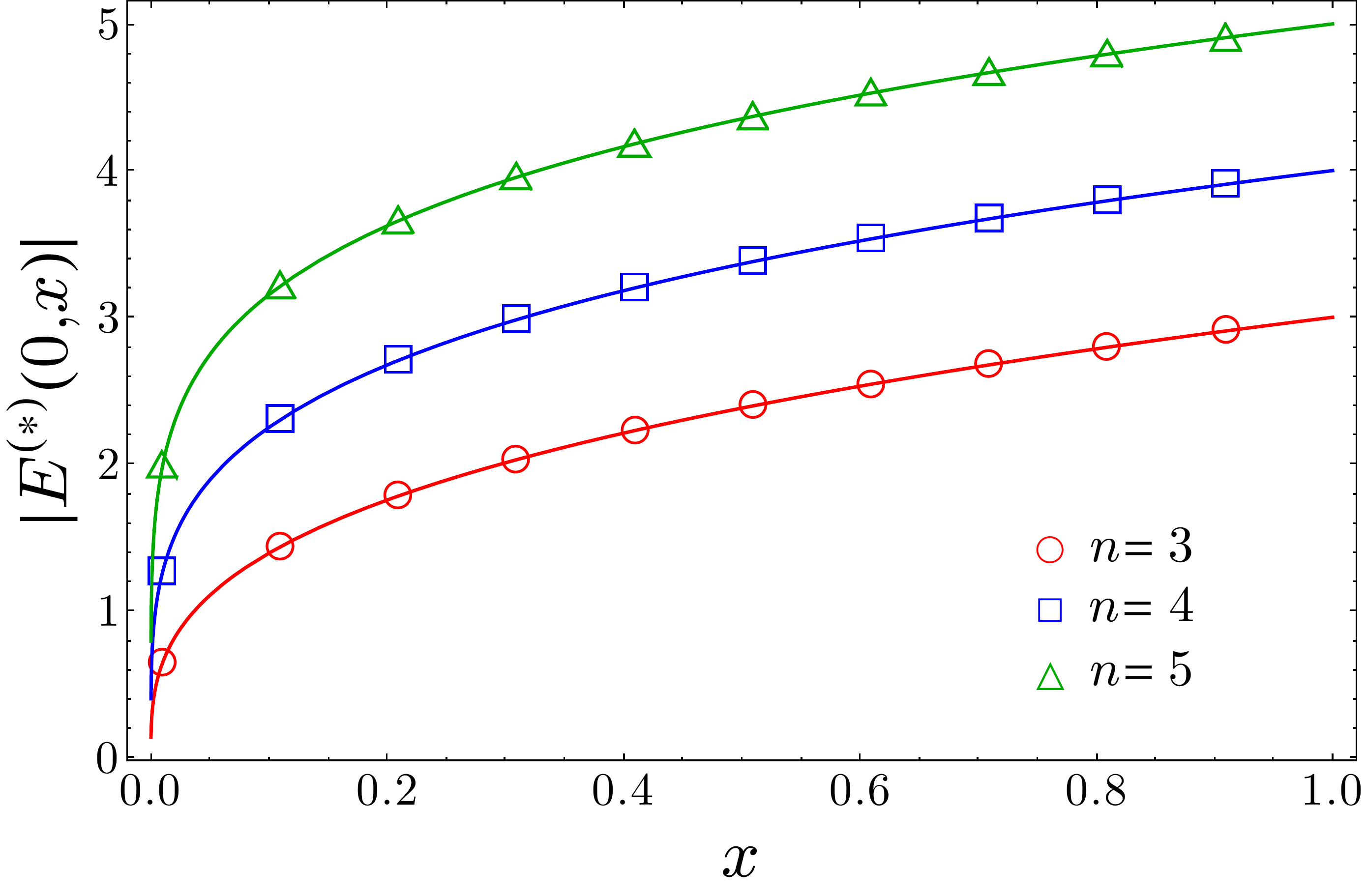}}
\subfigure
{\includegraphics[width=0.48\textwidth]{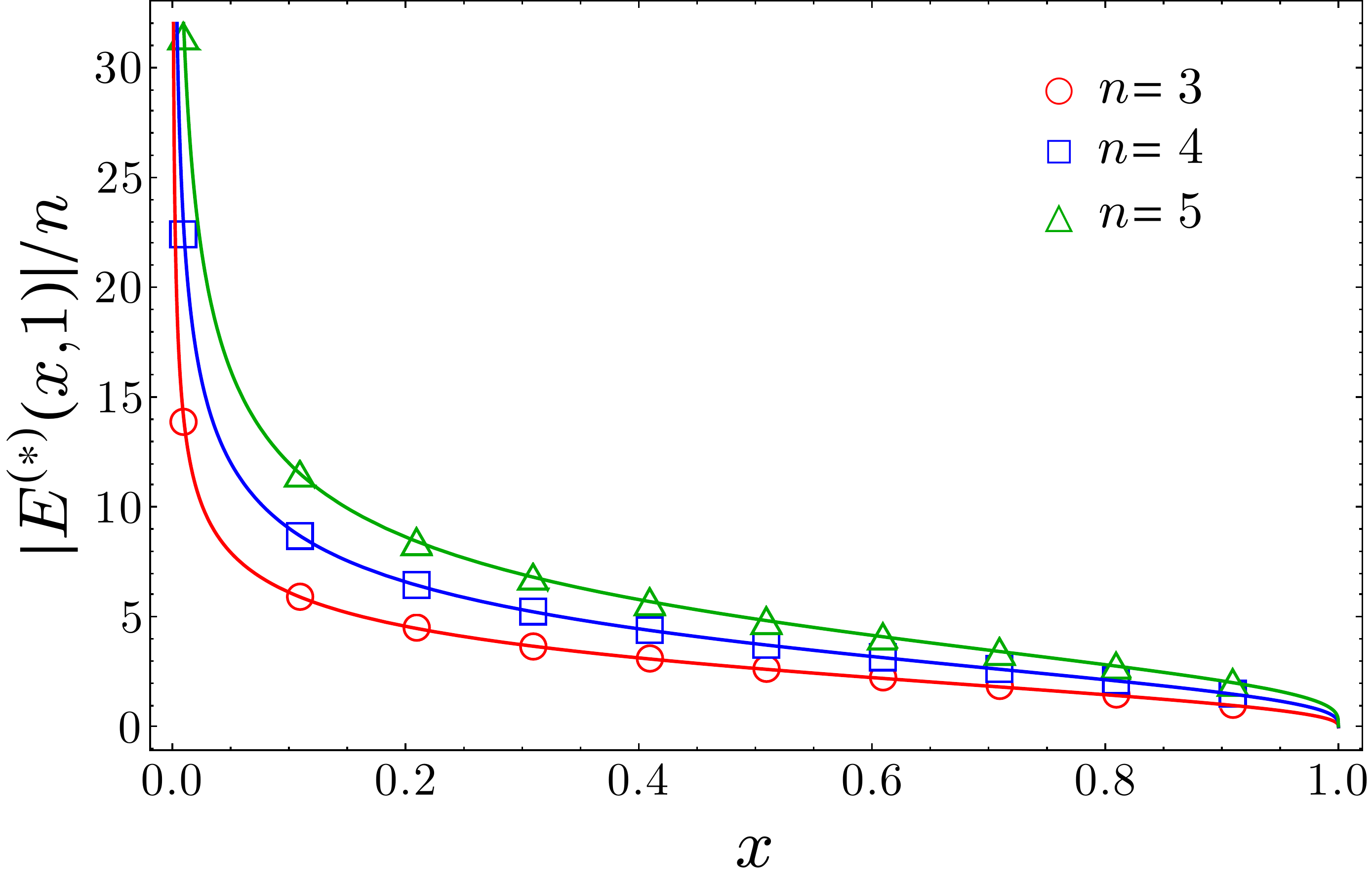}}
\subfigure
{\includegraphics[width=0.48\textwidth]{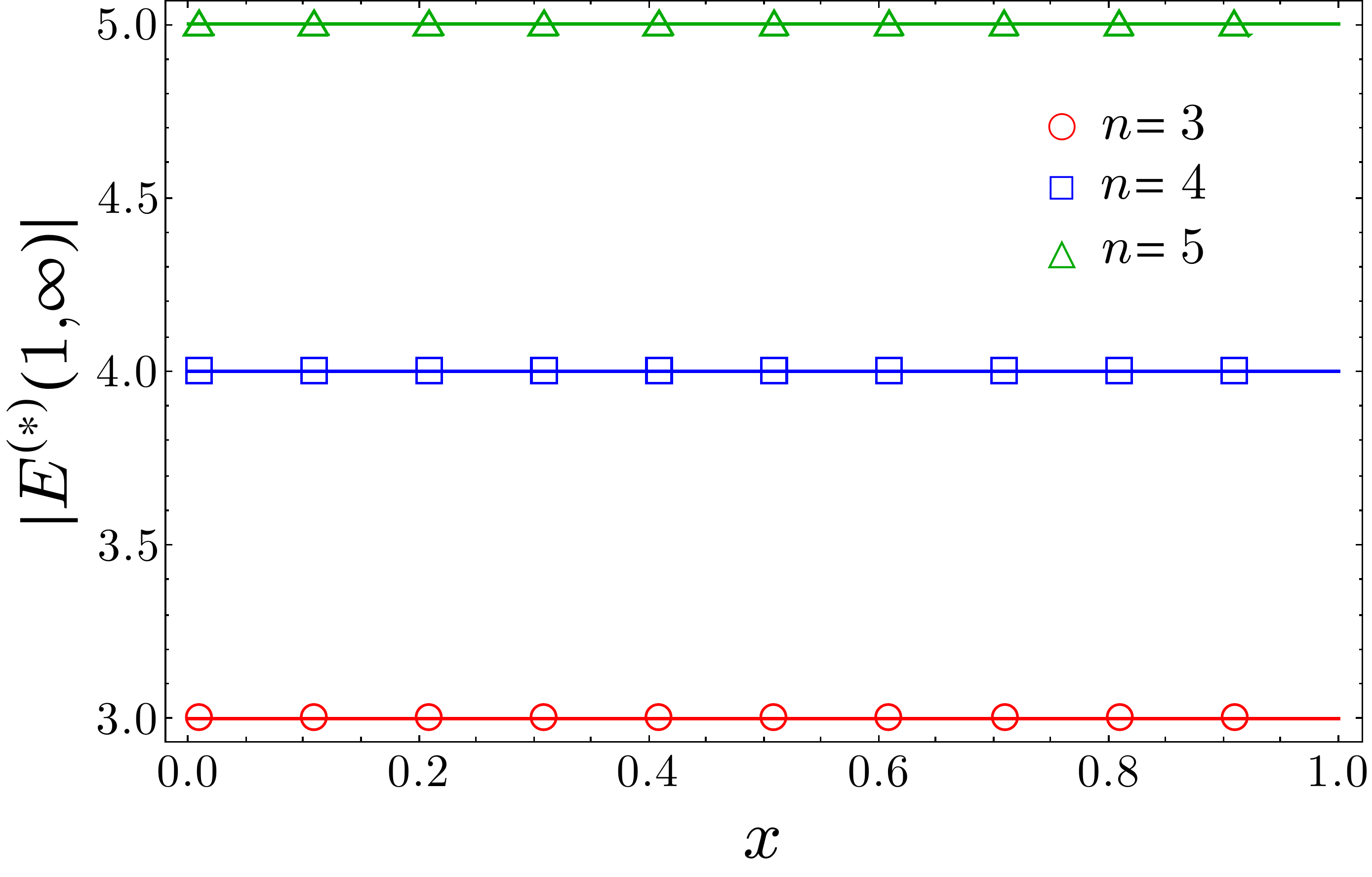}}
\subfigure
{\includegraphics[width=0.48\textwidth]{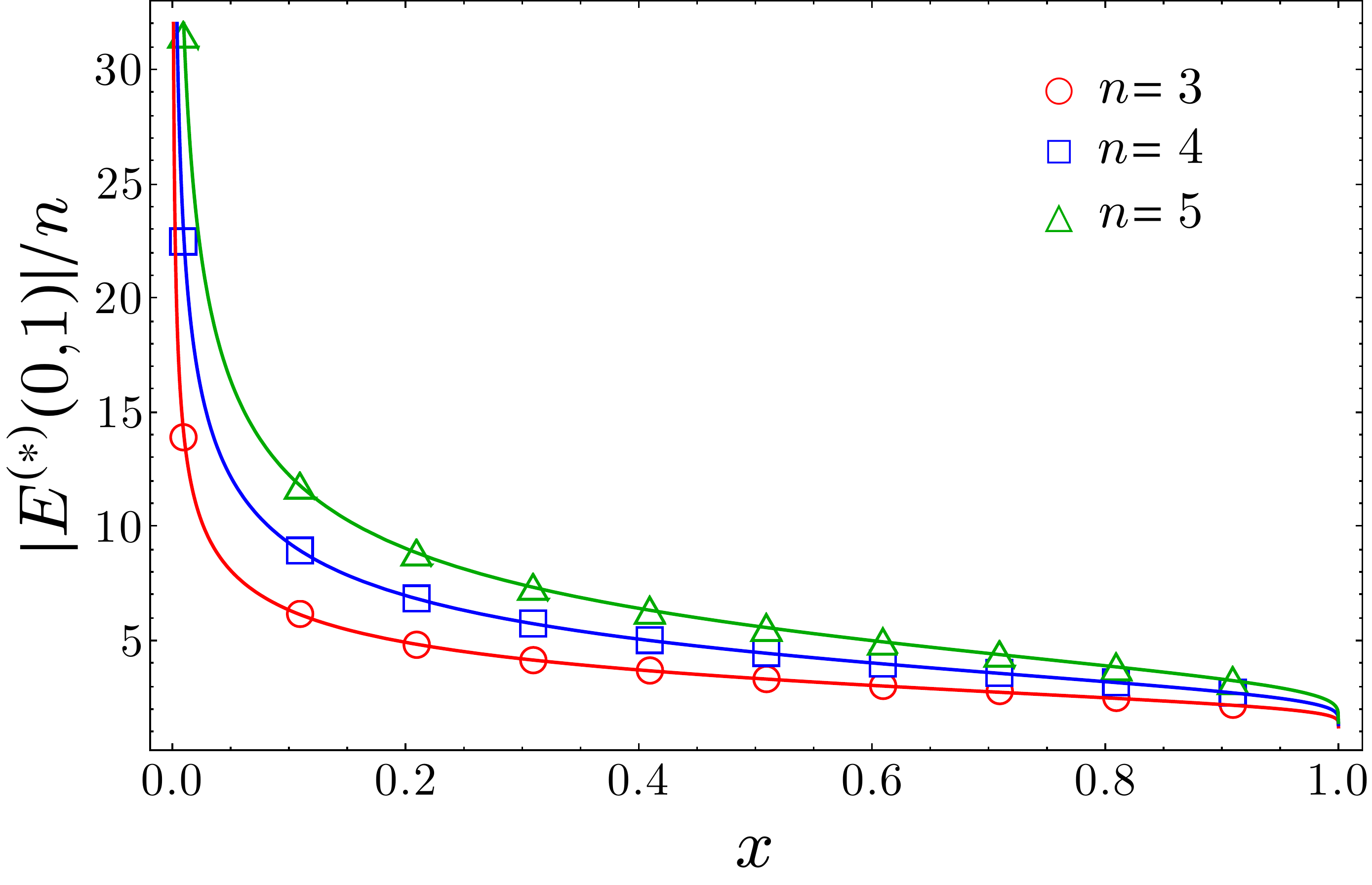}}
\caption{Numerical check of Eqs.~\eqref{eq:prime_form_id_1}-\eqref{eq:prime_form_id_4} involving
the regularised prime form $E^{(*)}(z, z')$. We test them for different values of the genus $n-1$. The points correspond to the 
direct numerical evaluation of the definition in Eq.~\eqref{eq:reg_prime_form} of $E^{(*)}(z, z')$. The continuous 
lines are the plot of the functions on the right hand side of 
Eqs.~\eqref{eq:prime_form_id_1}-\eqref{eq:prime_form_id_4}.}
\label{fig:prime_form_identities}
\end{figure}

The results for $n=2$ led us to conjecture the generalisation of Eqs.~\eqref{eq:prime_form_id_1}-\eqref{eq:prime_form_id_4} for higher genus. Unfortunately, we have not been able to 
find a proof for them, although they can be tested numerically with
great accuracy, as we show in Fig.~\ref{fig:prime_form_identities}.
The dots are the result of evaluating numerically the definition in Eq.~\eqref{eq:reg_prime_form} of $E^{(*)}(z,z')$, while the solid lines are the functions on the right hand side of the identities in Eqs.~\eqref{eq:prime_form_id_1}-\eqref{eq:prime_form_id_4}.

\end{document}